\documentclass[prd,showpacs,twocolumn,superscriptaddress,nofootinbib,floatfix,10pt]{revtex4-2}
\usepackage{setspace}
\usepackage{xcolor}
\usepackage[german,english]{babel}

\usepackage{bm}
\usepackage{braket}
\usepackage{slashed}
\usepackage{multirow}
\usepackage{epsfig}
\usepackage{epstopdf}
\usepackage{color}
\usepackage{bbm}
\usepackage{bbold}
\usepackage{soul}

\usepackage{amssymb}
\usepackage{amsmath}
\usepackage{colortbl}
\allowdisplaybreaks 
\newcommand{\be}{\begin{equation}}
\newcommand{\ee}{\end{equation}}
\newcommand{\bea}{\begin{eqnarray}}
\newcommand{\eea}{\end{eqnarray}}
\newcommand{\nn}{\nonumber}
\newcommand{\vep}{{\bm p}}

\newcommand{\vex}{{\bm x}}
\newcommand{\veX}{{\bm X}}
\newcommand{\velambda}{{\bm \lambda}}
\newcommand{\verho}{{\bm \rho}}
\newcommand{\Omfive}{O_5}
\newcommand{\ds}{\displaystyle}

\usepackage[
colorlinks=true,
linkcolor=blue,
breaklinks=true,
citecolor=blue]{hyperref}

\def\nn{\nonumber}

  \def\pa{\partial}   \def\dag{\dagger}

\newcommand{\spsi}{S}
\newcommand{\opsi}{O}

\newcommand{\itp}{\affiliation{CAS Key Laboratory of Theoretical Physics, Institute of Theoretical Physics,\\ Chinese Academy of Sciences, Beijing 100190, China}}

\newcommand{\ucas}{\affiliation{School of Physical Sciences, University of Chinese Academy of Sciences, Beijing 100049, China}}

\newcommand{\JSI}{\affiliation{Jozef Stefan Institute, Jamova 39, 1000 Ljubljana, Slovenia}}

\newcommand{\lis}{\affiliation{CeFEMA, Center of Physics and Engineering of Advanced Materials, Instituto Superior T{\'e}cnico, Avenida Rovisco Pais 1, 1049-001 Lisboa, Portugal}}

\graphicspath{{figs.dir/}}

\begin{document}

\title{Chromopolarizabilities of fully heavy baryons}

\author{Xiang-Kun Dong}\email{dongxiangkun@itp.ac.cn}
\itp\ucas

\author{Feng-Kun Guo}\email{fkguo@itp.ac.cn}
\itp \ucas

\author{Alexey Nefediev}\email{Alexey.Nefediev@ijs.si}
\JSI \lis

\author{Jaume Tarr\'us Castell\`a}
\email{jtarrus@iu.edu}
\affiliation{Department of Physics, Indiana University, Bloomington, Indiana 47401, USA}
\affiliation{Center for Exploration of Energy and Matter, Indiana University, Bloomington, Indiana 47408, USA}

\begin{abstract}
We compute the chromopolarizabilities of the fully heavy baryons $\Omega_{QQQ'}$ ($Q,Q'=b,c$) in the framework of potential nonrelativistic quantum chromodynamics. At leading order, the fully heavy hadrons are considered as ground chromo-Coulombic bound states. We find that the chromopolarizability $\beta_\Omega$ of a fully heavy baryon $QQQ$ is 2.6 times the chromopolarizability $\beta_\psi$ of the quarkonium $\bar{Q}Q$ with the same heavy quark flavor $Q$. This result is accurate up to the correction of the order 0.3 for $Q=b$ and provides an order-of-magnitude estimate for $Q=c$. We discuss the dependence of the ratio $\beta_\Omega/\beta_\psi$ on the heavy quark mass $m_Q$ and the strong coupling constant $\alpha_s$ as well as on the ratio of the masses $m_{Q'}/m_Q$, in the case not all quarks in the baryon are identical. Since the chromopolarizability of heavy hadrons defines the strength of their interaction at low energies mediated by soft gluons, which at long range hadronize into pairs of pions and kaons, our findings argue in favor of the existence of near-threshold states composed of pairs of fully heavy baryons.
\end{abstract}

\maketitle


\section{Introduction}\label{sec:int}

Light-meson exchanges are important as binding forces in nuclei and other hadrons as well as in extended objects like hadronic molecules. The Okubo-Zweig-Iizuka(OZI)-allowed light-meson exchanges are only possible between hadrons containing light quarks as constituents while fully heavy hadrons can interact and, in some cases, form bound states through soft-gluon exchanges. Theoretical foundations of such soft-gluon-driven interactions in heavy-quark systems were discussed in Ref.~\cite{Peskin:1979va}.
At the hadronic level, this mechanism corresponds to exchanges of pairs of light hadrons which are formally OZI suppressed, when the isospin or SU(3) breaking is switched off. A typical example of such a situation is provided by the 
double-$J/\psi$ system studied experimentally by the LHCb Collaboration~\cite{LHCb:2020bwg} and more recently by the ATLAS~\cite{ATLAS:jpsi} and CMS~\cite{CMS:jpsi} Collaborations. In particular, a theoretical coupled-channel analysis of the LHCb data for the $J/\psi J/\psi$ invariant mass distribution indicates the possible existence of a resonance near the double-$J/\psi$ threshold~\cite{Dong:2020nwy,Liang:2021fzr,Wang:2022jmb} 
with a large molecular component in its wave function~\cite{Dong:2020nwy,Liang:2021fzr}. Furthermore, it is argued in Ref.~\cite{Dong:2021lkh} that soft-gluon exchanges, hadronized in the form of correlated two-pion and two-kaon exchanges, might be strong enough to provide a sizable attraction between two $J/\psi$'s, consistent with the existence of a near-threshold pole on the first or second Riemann sheet of the energy complex plane. 

Then a natural question arises that whether or not similar mechanisms could be operative in the systems composed of two fully heavy baryons to result in the creation of dibaryon bound or virtual states. Interestingly, 
recent simulations using lattice quantum chromodynamics (QCD) indeed indicate that both double-$\Omega_{ccc}$ and 
double-$\Omega_{bbb}$ systems may be bound. In particular, the binding energy of the 
double-$\Omega_{ccc}$ system in the $^1S_0$ channel is computed by the HAL QCD Collaboration to be $-5.68^{+1.28}_{-0.90}$~MeV (with the electric Coulomb interaction ignored; the statistical and systematic errors have been added in quadrature)~\cite{Lyu:2021qsh} while the double-$\Omega_{bbb}$ system was found to be deeply bound in the $^1S_0$ channel, with a binding energy of $-89_{-12}^{+16}$~MeV~\cite{Mathur:2022nez}.
These results suggest that theoretical studies of the interactions in the systems formed by fully heavy hadrons are quite important and timely. 

Crucial information on such interactions is encoded in the chromopolarizability---the parameter which is determined by the intrinsic properties of a given fully heavy hadron and defines the strength of its interaction with soft gluons. 

The underlying idea of the approach stems from the fact that the effective field theory (EFT) Lagrangian for an $S$-wave heavy field $H$ in the external chromoelectric field $\bm{E}^a$ (with $a$ the color index) can be written in the form \cite{Luke:1992tm,Brambilla:2015rqa}
\begin{align}
L_{\rm EFT}^H=\int d^3X\,& H^\dag(t,\bm{X})\left\{i\pa_0+\frac{\bm{\nabla}_{\bm{X}}^2}{2m_H}\right.\nn\\
&\left.+\frac12\beta_H g^2\bm{E}^{a2}+\dots \right\}H(t,\bm{X}),
\label{LH}
\end{align}
where the scales $m_H$, $\alpha_s m_H$, and $\alpha_s^2 m_H$ (with $m_H$ for the heavy hadron mass and $\alpha_s$ for the strong coupling constant) are integrated out, so the ellipsis denotes higher-order operators, $g$ is the strong coupling, and the coefficient $\beta_H$ is the chromopolarizability mentioned above. If the size of a hadronic system is small compared with a typical length scale of the fluctuations in the nonperturbative QCD vacuum, then, to the leading order in the ratio of these scales, the interaction of the heavy hadron with the soft gluons can be considered as generated by two instantaneous color dipoles \cite{Peskin:1979va}. Formally, this picture is obtained by matching the Lagrangian in Eq.~\eqref{LH} with the one of weakly coupled potential nonrelativistic QCD (pNRQCD)~\cite{Pineda:1997bj,Brambilla:1999xf,Brambilla:2005yk}. In this way one can derive an analytical expression for the chromopolarizability; see, for example, Ref.~\cite{Brambilla:2015rqa} where such an approach was applied to the ground-state bottomonium. Equipped with the value of the chromopolarizability, one can use it as a building block to establish the strength of the interaction potential between the corresponding fully heavy hadrons.

The electromagnetic interaction between neutral composite particles is known as the van der Waals force. Two cases can be distinguished depending on the distance between the particles compared with their intrinsic scales~\cite{Brambilla:2017ffe}. On the one hand, the London potential arises if the time interval between the emission of the two photons is much larger than their travel time between the neutral particles. On the other hand, in the case of the Casimir-Polder interaction, the two photons are emitted almost simultaneously compared with their travel time. The interaction between two heavy quarkonia or two fully heavy baryons mediated by two-gluon exchanges can be viewed as a QCD analog of the van der Waals force. One such interaction is generated by the polarizability operator in the Lagrangian \eqref{LH} and corresponds to the Casimir-Polder type since the two gluons are emitted simultaneously. 
In most practical cases, the latter are nonperturbative, and it is necessary to consider their hadronization when constructing the van der Waals potential. The long-distance part of the potential is dominated by pairs of pions---the determination of the corresponding matrix elements of the operator $\bm{E}^{a2}$ can be found in Refs.~\cite{Novikov:1980fa,Pineda:2019mhw}. 
An analytic expression and plots of the two-pion exchange potential are provided in Ref.~\cite{Brambilla:2015rqa}.
At medium distances, the two-kaon contribution and the formation of the two-pion resonance, the $f_0(500)$, play a role. In this case, the potential can be constructed employing a dispersive technique~\cite{Dong:2021lkh}. Heavier resonances, quark exchanges and other short-range interactions are encoded in the contact term which renders
the resulting potential well-defined and provides cutoff-independent predictions, as required by the renormalization group analysis. Further discussions 
of the regularization procedure and a typical shape of the potential which arises can be found in Ref.~\cite{Dong:2021lkh}. In any case, the shape of the potential is independent of the value of the chromopolarizability, which enters as an overall multiplicative factor.

In this paper, we employ weakly coupled pNRQCD to calculate the chromopolarizability $\beta_\Omega$ of a fully heavy ground-state baryon $\Omega_{QQQ'}$ consisting of two heavy quarks of the same flavor $Q$ and mass $m_Q$ and the third quark of possibly (but not necessarily) a different flavor $Q'$ of the mass $m_{Q'}$. Therefore, for $Q,Q'=c,b$, we study the three-quark systems, $ccc$, $ccb$, $cbb$, and $bbb$, simultaneously. In particular, we obtain the relation between the chromopolarizabilities of a $QQQ'$ baryon, $\beta_\Omega$, and that of a $\bar{Q}Q$ meson, $\beta_\psi$, and establish the dependence of the ratio $\beta_\Omega/\beta_\psi$ on the ratio of the masses $m_{Q'}/m_Q$. We treat the heavy hadrons ($\bar{Q}Q$ meson or $QQQ'$ baryon) as purely Coulombic systems thus neglecting the nonperturbative dynamics inside of them.
This approximation is valid if the ratio of the nonperturbative and perturbative contributions to the energy of the system,
\be
\frac{E_{\rm np}}{E_{\rm pert}}\sim
\frac{\braket{\sigma r}}{\braket{\alpha_s/r}}
\sim \frac{\Lambda_{\rm QCD}^2}{\alpha_s^3m_Q^2},
\label{EnpEpert}
\ee
is small. Here
$\braket{\ldots}$ denotes averaging, the string tension $\sigma$, which introduces the nonperturbative scale related with confinement, was roughly estimated as $\sigma\sim\Lambda_{\rm QCD}^2$, and the mean size of the hadron was taken as {$\braket{r^{-1}}\sim \braket{r}^{-1}\sim\alpha_s m_Q$}, which is valid for a purely Coulombic system. For $\Lambda_{\rm QCD}=300$~MeV and $\alpha_s=0.35$ (see also Eq.~(\ref{eq:alphas}) below and the discussion around it) the ratio (\ref{EnpEpert}) takes values of the order unity and 0.1 for the $c$ and $b$ quarks, respectively. In other words, the approximation of a fully heavy hadron by a purely color-Coulombic system predictably works well for the ground-state meson or baryon composed of the bottom quarks. In the meantime, corrections may appear comparable with the leading term for the ground-state charmonium and $\Omega_{ccc}$ baryon. In the latter case we are aimed at an order-of-magnitude estimate.

The paper is organized as follows. In Sec.~\ref{sec:betaofbb} we consider a heavy quarkonium and present a numerical computation of its chromopolarizability based on placing the system in a finite box. We compare our results with the analytical ones contained in the literature and find a good agreement. Therefore, equipped with the investigation method, we proceed to Sec.~\ref{sec:pola} and evaluate the chromopolarizability of a fully heavy baryon. We conclude in Sec.~\ref{sec:concl}. Various details related to the calculations performed in this paper are collected in appendices. In particular, generalized Jacobi coordinates for a three-body system are introduced in Appendix~\ref{app:Jacobi}; in Appendix~\ref{app:details} we provide some details of numerical calculations of the fully heavy baryon chromopolarizability; Appendix~\ref{app:SE2D} is devoted to calculations for a three-quark system in a finite box; finally, in Appendix~\ref{Omixing} we discuss the effect of the mixing of the octet representations for a fully heavy baryon.
\newpage

\section{Chromopolarizability of a heavy quarkonium}\label{sec:betaofbb}

\subsection{Derivation of $\beta_\psi$}
\label{sec:betapsi}

In this section, as a warm-up and to introduce the necessary essentials, we reproduce the results for the chromopolarizability of a heavy quarkonium $\bar{Q}Q$. We follow the lines of Ref.~\cite{Brambilla:2015rqa}. The underlying idea of the approach is a multipole expansion performed to the order ${\cal O}(r)$ (with $r$ for the interquark separation) in the nonrelativistic Lagrangian for the given heavy hadronic system. Since the $Q\bar{Q}$ pair can be either in the color-singlet or color-octet state, 
the Lagrangian is written in terms of the effective fields $\spsi$ (for the singlet) and $\opsi$ (for the octet). Up to leading and next-to-leading order in the heavy quark mass and multipole expansion the Lagrangian of the interacting singlet and octet fields takes the form \cite{Pineda:1997bj,Brambilla:1999xf} 
\begin{align}
&\mathcal{L}_{\rm pNRQCD}^{(0)}=\int d^3r\text{Tr}\left[\spsi^\dagger(i\partial_0-\hat{h}_S)\spsi\right.
\label{LQQ1}
\\
&+\opsi^{a\dagger}(i\partial_0-\hat{h}_O)\opsi^a
+(\spsi^{\dagger}\bm{r}\cdot g\bm{E}^a \opsi^a+H.c.)\Big],\nn
\end{align}
where the potential which appears in front of the singlet--octet mixing term has been set to unity at the given level of matching. Further details on pNRQCD can be found in the reviews~\cite{Brambilla:2004jw, Pineda:2011dg}.

In the Lagrangian (\ref{LQQ1}) $\hat{h}_{\spsi}$ and $\hat{h}_{\opsi}$ are the singlet and octet Hamiltonians, respectively,
\begin{align}
\hat{h}_{\spsi}&=\hat{T}({\bm p})+V_{\spsi}(r)=-\frac{1}{m_Q}\bm{\nabla}_r^2-\frac43\frac{\alpha_s}{r},
\label{hSpsi}\\
\hat{h}_{\opsi}&=\hat{T}({\bm p})+V_{\opsi}(r)=-\frac{1}{m_Q}\bm{\nabla}_r^2+\frac16\frac{\alpha_s}{r},
\label{hOpsi}
\end{align}
where $m_Q$ is the mass of the quark and only the terms responsible for the relative motion in the system are retained.

The heavy quarkonium $\bar{Q}Q$ (for brevity we denote it as $\psi$) of a mass $M_\psi$ is identified with the ground state of the Hamiltonian (\ref{hSpsi}),
\be
\hat{h}_{\spsi}\ket{\psi}=E_\psi\ket{\psi},
\ee
with $E_\psi=M_\psi-2m_Q$ for the binding energy. The wave function $\ket{\psi}$ in the coordinate space can be decomposed into the radial and angular parts,
\be
\braket{\bm{r}|\psi}=\frac{1}{r}u(r)Y_{00}(\hat{r}),
\ee
where the radial wave function $u(r)$ is normalized as
\be
\int_0^\infty dr|u(r)|^2=1
\ee
and obeys the {radial Schr{\"o}dinger} equation
\be
\left(-\frac{1}{m_Q}\frac{\partial^2}{\partial r^2}-\frac43\frac{\alpha_s}{r}\right)u(r)=E_\psi u(r).
\ee

At the same time, the spectrum of the octet Hamiltonian $\hat{h}_O$ from Eq.~(\ref{hOpsi}) consists of the continuum states $\ket{p,l,l_z}$ such that
\be
\hat{h}_{\opsi}\ket{p,l,l_z}=E_p\ket{p,l,l_z},
\ee
where $p$ is the 3-momentum while $l$ and $l_z$ are the orbital angular momentum and its projection, respectively. In the coordinate space one has
\be
\braket{\bm{r}|p,l,l_z}=\frac{1}{r}v_{p,l}(r)Y_{ll_z}(\hat{r}),
\ee
with the radial wave function $v_{p,l}(r)$ 
normalized as
\be
\int_0^\infty dr|v_{p,l}(r)|^2=1
\ee
and obeying the eigenstate equation
\be
\left(-\frac{1}{m_Q}\frac{\partial^2}{\partial r^2}+\frac{l(l+1)}{m_Q r^2}+\frac16\frac{\alpha_s}{r}\right)v_{p,l}(r)=E_p v_{p,l}(r).
\ee

Unlike Ref.~\cite{Brambilla:2015rqa} where the exact Coulombic eigenfunction $\braket{{\bm r}|\vep,1}$ was used, in this work we place the system in a finite box of the size $L_{\rm box}$ in the radial direction and impose the boundary conditions 
\be
u(0)=u(L_{\rm box})=0,\quad v_{p,l}(0)=v_{p,l}(L_{\rm box})=0.
\label{ubound}
\ee

Matching weakly coupled pNRQCD quoted in Eq.~\eqref{LQQ1} to the EFT defined by the Lagrangian in Eq.~\eqref{LH}, with $H=\psi$, one can obtain the chromopolarizability as
\be
\beta_\psi=\frac19\bra{\psi} \bm{r}\frac{1}{\hat{h}_{\opsi}-E_\psi}\bm{r}\ket{\psi}.
\label{betapsi}
\ee
This expression can be understood as a double emission of soft gluons with an octet state propagating between the two emission vertices. 

Before we proceed to the numerical calculations of the chromopolarizability $\beta_\psi$, we notice that Eq.~(\ref{betapsi}) allows one to make some general conclusions about the behavior of $\beta_\psi$ as a function of $m_Q$ and $\alpha_s$. Indeed, in the Coulombic system at hand the following simple relations hold (as before, $\braket{\ldots}$ stand for the averaged values):
\be
\braket{p}\sim \frac1{\braket{r}}\sim\alpha_s m_Q,\quad E_\psi\sim \alpha_s^2m_Q,
\label{scales}
\ee
which imply that
\be
\beta_\psi=\frac{C_\psi}{\alpha_s^4m_Q^3},
\label{Cpsidef}
\ee
with $C_\psi$ a constant independent of $\alpha_s$ and $m_Q$. The expression for $\beta_\psi$ quoted in Eq.~(\ref{Cpsidef}) was obtained analytically in Ref.~\cite{Brambilla:2015rqa}, and the factor $C_\psi$ was evaluated to be $0.93$.

Since the chromopolarizability demonstrates such a strong dependence on $m_Q$ and $\alpha_s$, its numerical value depends significantly on the renormalization scale and scheme used to obtain these two quantities. In principle, the physical observables should be independent of the renormalization scale and scheme used, however, working in perturbation theory some dependence is unavoidable. In particular, the expressions for the chromopolarizability used in this work are derived 
at leading order, so a strong dependence on the renormalization scale has to be anticipated. Therefore, to sidestep this issue, we provide the results for the constant $C_\psi$ and quote {some} representative values of the chromopolarizability $\beta_\psi$ in Table~\ref{tab:betaspsi} below.

\subsection{Numerical evaluation of ${\beta_\psi}$}
\label{sec:numpsi}

Here we recalculate the factor $C_\psi$ by putting the system in a finite box; see Eq.~(\ref{ubound}). The same technique will be applied later to the fully heavy baryon case where analytical expressions are not available.

In order to proceed with the numerical calculations, we use the completeness condition for the continuum spectrum to write
\begin{align}
\beta_\psi&=\frac19\sum_{p,l,l_z}\sum_{r_i=x,y,z}\bra{\psi}r_i\ket{p,l,l_z}\frac{1}{E_p-E_\psi}\bra{p,l,l_z}r_i\ket{\psi}\nn\\
&=\frac19\sum_{p}\frac{|I^{(r)}(p)|^2}{E_p-E_\psi},
\label{eq:betabb}
\end{align}
where the integration over the momentum is transformed to a summation as the system is put in a finite box. The function $I^{(r)}(p)$ is defined as
\be
I^{(r)}(p)\equiv \int_0^\infty rdr\, u(r)v_{p,1}(r),
\label{Ir}
\ee
where only the term with $l=1$ contributes since the octet field is a $P$-wave operator, and the following easily verifiable matrix elements were used in the calculation: 
\begin{align*}
\braket{\psi|x|p,l,l_z}&=\mp\delta_{l,1}\delta_{l_z,\pm1}\frac{1}{\sqrt6}I^{(r)}(p),\\
\braket{\psi|y|p,l,l_z}&=-\delta_{l,1}\delta_{l_z,\pm1}\frac{i}{\sqrt6}I^{(r)}(p),\\
\braket{\psi|z|p,l,l_z}&=\delta_{l,1}\delta_{l_z,0}\frac1{\sqrt3}I^{(r)}(p).
\end{align*}

A direct numerical computation performed according to Eqs.~(\ref{eq:betabb}) and (\ref{Ir}) demonstrates a perfect agreement of the result obtained with the scaling behavior described in Eq.~(\ref{Cpsidef}) with
\be
C_\psi\approx 0.93.
\label{Cpsi}
\ee
This result applies both to charmonium ($\bar{c}c$) and bottomonium ($\bar{b}b$) systems. Moreover, the value in Eq.~\eqref{Cpsi} is in a good numerical agreement with the result for the chromopolarizability of a bottomonium reported in Ref.~\cite{Brambilla:2015rqa}.

Before we come to a numerical evaluation of the chromopolarizability $\beta_\psi$, let us compare the exact result (\ref{Cpsi}) with a simple estimate which will also be convenient for the discussions of the baryons below. To this end we 
notice that, in the Coulombic system at hand, the energies of the bound states are negative while the eigenenergies in the continuum spectrum are positive. Therefore, 
reconstructing the identity from the closure relation and then
setting $E_p=0$ in Eq.~(\ref{eq:betabb}), we arrive at an upper bound, 
\be
\beta_\psi\lesssim \frac{\braket{r^2}}{9|E_\psi|},
\label{betapsiest1}
\ee
where $\braket{\ldots}$ stands for averaging over the ground state $\ket{\psi}$ of the Hamiltonian (\ref{hSpsi}). Then, with the help of the easily verifiable relations,
\be
E_\psi=-\frac{4}{9}\alpha_s^2m_Q,\quad 
\braket{r^2}=\frac{27}{4(m_Q\alpha_s)^2},
\ee
we finally arrive at the inequality
\be
C_\psi\lesssim\frac{27}{16}\approx 1.7.
\label{Cpsi1}
\ee

The exact result (\ref{Cpsi}) complies well with the upper bound (\ref{Cpsi1}). 

Notice that Eq.~(\ref{betapsi}) can be approximately rewritten in the form
\be
\beta_\psi\approx\frac19\frac{\braket{ r^2}}{\braket{\hat{T}({\bm p})}+\braket{V_{\opsi}(r)}-E_\psi},
\label{betapsi2}
\ee
where $\braket{V_{\opsi}(r)}=\alpha_s^2m_Q/9$. Then it is easy to check that the exact result (\ref{Cpsi}) is rather accurately reproduced for the averaged kinetic energy $\braket{\hat T({\bm p})}\approx |E_\psi|/2$.

Let us now estimate the correction to the chromopolarizability $\beta_\psi$ due to a possible contribution of the nonperturbative interaction in the system. To this end we treat the confining interaction $\sigma r$ as a perturbation to arrive at the correction,
\be
\delta\beta_\psi=\frac19\bra{\psi} \bm{r}\frac{1}{\hat{h}_{\opsi}-E_\psi}\sigma r
\frac{1}{\hat{h}_{\opsi}-E_\psi}\bm{r}\ket{\psi},
\label{dbetapsi}
\ee
that allows one to find the ratio,
\be
\frac{\delta\beta_\psi}{\beta_\psi}
\sim \frac{\Lambda_{\rm QCD}^2}{\alpha_s^3 m_Q^2}.
\label{dbb}
\ee

Quite naturally, the ratio (\ref{dbb}) is defined by the same combination of the scales involved as provided in Eq.~(\ref{EnpEpert}), so that the numerical estimates made after Eq.~(\ref{EnpEpert}) are valid here as well. Clearly, the same conclusion holds for the ground-state fully heavy baryons to be studied below. Thus, for the chromopolarizabilities of the ground-state charmonium and $\Omega_{ccc}$ baryon we pretend to provide order-of-magnitude estimates which nevertheless are expected to lie in the right ballpark, especially given the large uncertainties they have (see the discussion below). On the other hand, the corrections to the chromopolarizability of the ground-state bottomonium and $\Omega_{bbb}$ baryon due to the nonperturbative dynamics are expected to be at the level of about 10\%.

Now, to provide numerical estimates for the chromopolarizabilities of the ground-state $\bar{b}b$ and $\bar{c}c$ quarkonia, we use the following values of the heavy quark masses:
\begin{align}
&m_c^{\rm RS'}(1~{\rm GeV})=1.496(41)~\mbox{GeV},\nn\\[-3mm] 
\label{eq:masses}\\[-3mm]
&m_b^{\rm RS'}(1~{\rm GeV})=4.885(41)~\mbox{GeV},\nn
\end{align}
obtained in the renormalon-subtracted scheme (RS$^\prime$) of Ref.~\cite{Peset:2018ria} which improves the convergence of the perturbative expansion while keeping the leading order potential unchanged. The strong coupling constant is taken at the renormalization scale $\nu_r=1.5$~GeV, which is large enough for a reasonable convergence of perturbation theory while minimizing contributions of logarithms associated with the soft scale. Using the \texttt{RunDec} routine at the $4$-loop accuracy~\cite{Chetyrkin:2000yt,Herren:2017osy} we find
\begin{align}
&\alpha_s(\nu_r=1.5~{\rm GeV{\rm}})=0.3485.
\label{eq:alphas}
\end{align}
The uncertainty of the chromopolarizabilities associated to picking the renormalization scale is estimated by varying $\nu_r$ {between} $1$ and $2$~GeV, which corresponds to the following boundary values of the strong coupling constant:
\begin{align}
&\alpha_s(\nu_r=1~{\rm GeV{\rm}})=0.4798,\nn\\[-2mm] 
\label{eq:alphas2}\\[-2mm]
&\alpha_s(\nu_r=2~{\rm GeV{\rm}})=0.3015.\nn
\end{align}
The results for the mean radii and chromopolarizabilities of the heavy $\bar{Q}Q$ mesons with $Q=c,b$ are listed in Table~\ref{tab:betaspsi}. The large uncertainties, especially for $\beta_\psi$, stem from a large spread in the values of $\alpha_s$ quoted in Eq.~\eqref{eq:alphas} and \eqref{eq:alphas2} and the $\alpha_s^{-4}$ scaling of $\beta_\psi$ in Eq.~\eqref{Cpsidef}. 

\begin{table}[t]
\caption{The mean radii and chromopolarizabilities of the ground-state heavy $\bar{Q}Q$ mesons with $Q=c,b$ evaluated in this work. See the main text for the discussion of the uncertainties.}
\label{tab:betaspsi}
\begin{ruledtabular}
\renewcommand\arraystretch{1.7}
\begin{tabular}{lcc}
State & $\bar{c}c$ & $\bar{b}b$ \\ \hline
$\langle r\rangle$ [fm] & $0.85_{-0.23}^{+0.13}$ & $0.26_{-0.07}^{+0.04}$ \\
\hline
$\beta_\psi$ [GeV$^{-3}$] & $19_{-14}^{+15}$ & $0.54_{-0.39}^{+0.43}$ 
\end{tabular}
\end{ruledtabular}

\end{table}

\section{Chromopolarizability of a fully heavy baryon}
\label{sec:pola}

\subsection{Derivation of $\beta_\Omega$}

As was discussed in the Introduction, we consider a baryon made of two quarks of the same flavor (with the mass $m_Q$) and the third quark of potentially another flavor (with the mass $m_{Q'}$).
The interactions between three heavy quarks in an EFT context incorporating the heavy quark and multipole expansions have been worked out in Ref.~\cite{Brambilla:2005yk}. 

The three heavy quark fields can be decomposed into a singlet ($S$), two octets ($O^A$, $O^S$) and a decuplet ($\Delta$) in the color space,
\be
3\otimes 3\otimes 3= 1\oplus 8\oplus 8\oplus 10,
\ee
however, only the octet fields can couple to the singlet via a single chromoelectric field insertion. These fields depend on the three Jacobi coordinates. The choice of the latter is not unique, and for the future convenience we stick to the
following relations (see Appendix~\ref{app:Jacobi} for further details on the Jacobi coordinates in a three-body system):
\begin{align}
\bm{X}&=\frac{m_Q(\bm{x}_1+\bm{x}_2)+m_{Q'}\bm{x}_3}{M},&\nonumber\\ 
\bm{\lambda}&=\frac{2}{\zeta}\left(\frac{\bm{x}_1+\bm{x}_2}{2}-\bm{x}_3\right),&\label{eq:Jacobi}\\
\bm{\rho}&=\bm{x}_1-\bm{x}_2,&\nonumber 
\end{align}
where $\bm{x}_1$, $\bm{x}_2$, and $\bm{x}_3$ are the positions of the quarks {with the masses $m_Q$, $m_Q$, and $m_{Q'}$, respectively}, $M=2m_Q+m_{Q'}$, and 
\begin{equation}
 \zeta=\sqrt{M/m_{Q'}}.
\end{equation}
The case of $m_{Q'}=m_Q$ corresponds to $\zeta=\sqrt{3}$. Then
\be
\frac12m_Q\left(\dot{\bm{x}}_1^2+\dot{\bm{x}}_2^2\right)+\frac12m_{Q'}\dot{\bm{x}}_3^2=\frac12M\dot{\bm{X}}^2+\frac14m_Q\left(\dot{\bm{\lambda}}^2+\dot{\bm{\rho}}^2\right)
\label{lagkin0}
\ee
and, therefore, the kinetic energy of the relative motion in the baryon is
\be
\hat{T}(\vep_\rho,\vep_\lambda)=\frac{\hat{\vep}_\rho^2+\hat{\vep}_\lambda^2}{m_Q}=-\frac{1}{m_Q}\left(\bm{\nabla}_{\rho}^2+\bm{\nabla}_{\lambda}^2\right),
\ee
where the hat stands for a differential operator in the coordinate space, with the eigenenergy 
\be
E_{p_\rho,q_\lambda}=\frac{p_\rho^2+p_\lambda^2}{m_Q}.
\label{Epq}
\ee

The Hamiltonian in a given color representation (${\cal R}\in\{S,O^S,O^A\}$) reads
\be
\hat{h}_{\cal R}=\hat{T}(\vep_\rho,\vep_\lambda)+V_{\cal R}({\bm\rho},{\bm\lambda}),
\label{hR}
\ee
with
\begin{align}
V_S&=-\frac{2\alpha_s}{3}\left(\frac{1}{|\bm{\rho}|}+\frac{2}{|\bm{\rho}+\zeta\bm{\lambda}|}+\frac{2}{|\bm{\rho}-\zeta\bm{\lambda}|}\right),
\label{hS}\\
V_{O^S}&=\frac{\alpha_s}{6}\left(\frac{2}{|\bm{\rho}|}-\frac{5}{|\bm{\rho}+\zeta\bm{\lambda}|}-\frac{5}{|\bm{\rho}-\zeta\bm{\lambda}|}\right),
\label{hOS}\\
V_{O^A}&=-\frac{\alpha_s}{6}\left(\frac{4}{|\bm{\rho}|}-\frac{1}{|\bm{\rho}+\zeta\bm{\lambda}|}-\frac{1}{|\bm{\rho}-\zeta\bm{\lambda}|}\right).
\label{hOA}
\end{align}

For the chromopolarizability of the baryon $\Omega_{QQQ'}$ we consider the dipolar couplings of the singlet with other fields. In particular, there are two such couplings to the octet fields,
\begin{align}
{\cal L}_{\rm pNRQCD}^{\mbox{\scriptsize $S$-$O$}}&=\int d^3\rho\, d^3\lambda \left\{\frac{1}{2\sqrt{2}}\left[S^{\dag}\bm{\rho}\cdot g\bm{E}^a O^{Sa}+H.c.\right]\right.\nonumber\\
&-\left.\frac{\zeta}{2\sqrt{6}}\left[S^\dag \bm{\lambda}\cdot g\bm{E}^aO^{Aa}+H.c.\right]\right\}\,,\label{dlg}
\end{align}
where, similar to the case of heavy quarkonium, the potentials which appear as the coefficients in front of each term in Eq.~(\ref{dlg}) are set to unities at the tree level of matching.

Strictly speaking, the octet fields $O^S$ and $O^A$ can mix~\cite{Brambilla:2013vx}, however the effect of such mixing on the chromopolarizability is negligibly small, so we disregard it here and check the validity of this neglect \emph{a posteriori}---see Appendix~\ref{Omixing} for the details. The (small) effect of mixing is then treated as a source of the systematic uncertainty. 

We can now define the chromopolarizability $\beta_\Omega$ in the same way as it was done for the quarkonium, with the only difference that now there are two terms corresponding to two different dipoles in Eq.~\eqref{dlg}. We, therefore, proceed along the lines of Ref.~\cite{Brambilla:2015rqa} and define a lower-energy EFT for the ground state $\Omega_{QQQ'}$ interacting with gluonic fields assuming that the typical energies are smaller than the binding energy of the $\Omega_{QQQ'}$ baryon. Then the Lagrangian again takes the same form as given in Eq.~(\ref{LH}), with $H=\Omega$, $m_{\Omega}=M+E_\Omega$, and $E_\Omega$ for the ground-state binding energy. 

Thus, one can derive an explicit expression for the chromopolarizability $\beta_\Omega$ by matching Eqs.~\eqref{dlg} and \eqref{LH},
\be
\beta_\Omega=\beta_\Omega^{(\rho)}+\beta_\Omega^{(\lambda)},
\ee
where
\be
\begin{split}
\beta_\Omega^{(\rho)}&=\frac{1}{12}\bra{\Omega}\bm{\rho}\frac{1}{\hat{h}_{O^S}-E_\Omega}\bm{\rho}\ket{\Omega},\\
\beta_\Omega^{(\lambda)}&=
\frac{\zeta^2}{36}\langle \Omega|\bm{\lambda}\frac{1}{\hat{h}_{O^A}-E_\Omega}\bm{\lambda}|\Omega\rangle.\label{pol}
\end{split}
\ee

The ground state $|\Omega\rangle$ is defined through the Schr{\"o}dinger equation
\begin{align}
\hat{h}_S|\Omega\rangle=E_\Omega|\Omega\rangle.\label{eq:SEb}
\end{align}

\subsection{Evaluation of $\beta_\Omega$}
\label{sec:betaomega}

\subsubsection{Some generalities}

Consider first the matrix element 
\be
\langle \Omega|\bm{\rho}\frac{1}{\hat{h}_{O^S}-E_\Omega}\bm{\rho}|\Omega\rangle,
\label{Mrho}
\ee
where the operator $\hat{h}_{O_S}$ is given by Eqs.~(\ref{hR}) and (\ref{hOS}). To proceed we employ the completeness of the continuum eigenstates of the operator $\hat{h}_{O^S}$,
\be
\hat{h}_{O^S}|\Psi_\nu^S\rangle=E_\nu^S|\Psi_\nu^S\rangle,
\ee
where the quantum number $\nu$ includes both the discrete angular momenta and the continuous quantum numbers, that is, momenta; see Eq.~(\ref{Psic}) below for details. 

Then, for the matrix element (\ref{Mrho}) we find
\bea
&& \langle \Omega|\bm{\rho}\frac{1}{\hat{h}_{O^S}-E_\Omega}\bm{\rho}|\Omega\rangle
=3\langle \Omega|\rho_z\frac{1}{\hat{h}_{O^S}-E_\Omega}\rho_z|\Omega\rangle\nonumber\\
&& =3\sum_{\nu,\nu'}\langle \Omega|\rho_z|\Psi_\nu^S\rangle\langle \Psi_\nu^S|\frac{1}{\hat{h}_{O^S}-E_\Omega}|\Psi_{\nu'}^S\rangle\langle \Psi_{\nu'}^S|\rho_z|\Omega\rangle\nn\\
&&=3\sum_\nu\frac{|\langle \Omega|\rho_z|\Psi_\nu^S\rangle|^2}{E_\nu^S-E_\Omega},
\label{matrel1}
\eea
where we have used that the $x$, $y$, and $z$ directions contribute equally. Applying the same approach to the matrix element
\be
\langle \Omega|\bm{\lambda}\frac{1}{\hat{h}_{O^A}-E_\Omega}\bm{\lambda}|\Omega\rangle,
\ee
we finally arrive at the following expressions for the two contributions to the chromopolarizability $\beta_\Omega$:
\be
\begin{split}
\beta_\Omega^{(\rho)}&=\frac14
\sum_\nu\frac{|\langle \Omega|\rho_z|\Psi_\nu^S\rangle|^2}{E_\nu^S-E_\Omega},
\\
\beta_\Omega^{(\lambda)}&=
\frac{\zeta^2}{12}
\sum_\nu\frac{|\langle \Omega|\lambda_z|\Psi_\nu^A\rangle|^2}{E_\nu^A-E_\Omega},
\label{pol2}
\end{split}
\ee
where $E_\nu^A$ and $\Psi_\nu^A$ are the eigenenergy and the corresponding eigenfunction of the Hamiltonian $\hat{h}_{O^A}$,
\be
\hat{h}_{O^A}|\Psi_\nu^A\rangle=E_\nu^A|\Psi_\nu^A\rangle.
\ee

\subsubsection{The singlet wave function}

To proceed with the bound state spectrum of a heavy baryon we introduce a hyperspherical basis~\cite{FabredelaRipelle:1988zz,Hasenfratz:1980ka},
\be
\rho=R\cos\theta,\quad\lambda=R\sin\theta,
\label{Rthetadef}
\ee
so that a complete 6-dimensional set of variables in the coordinate space reads
\be
\{R,\theta,\hat{\bm \rho},\hat{\bm \lambda}\}\equiv\{R,\Omfive\},
\ee
where $\hat{\bm \rho}$ and $\hat{\bm \lambda}$ are the unit vectors in the directions of the Jacobi coordinates $\verho$ and $\velambda$, respectively, and
\be
\int d\Omfive=\int_0^{\pi/2}\sin^2\theta\cos^2\theta\, d\theta\int d\hat{\bm \rho}\int d\hat{\bm \lambda}=\pi^3.
\ee

Since the spin variables are factorized out, they are not considered here. Then the bound-state wave function can be decomposed in a set of $K$-harmonics \cite{Simonov:1965ei,Badalian:1966wm,Calogero:1968zz},
\be
\Psi^{(b)}(R,\Omfive)=\frac{1}{R^{5/2}}\sum_{K,\alpha}\psi_{K,\alpha}^{(b)}(R){\cal Y}_{K,\alpha}(\Omfive),
\label{Psi}
\ee
where $\psi^{(b)}$ and ${\cal Y}$ are the radial and angular parts of the wave function. The latter are known as the hyperspherical harmonics, and their explicit form can be built through spherical functions and Jacobi polynomials; see, for example, Ref.~\cite{Ballot:1979hd}. {Here}
\be
\alpha\equiv \{L,L_z,l_\rho,l_\lambda\},
\ee
with $l_\rho$, $l_\lambda$, and $L$ for the angular momenta ($L$ being the total one). {Then} $K=2n_r+l_\rho+l_\lambda$ is a non-negative integer number and $n_r$ is the radial excitation quantum number. The wave functions of the form (\ref{Psi}) are normalized as
\be
\langle \Psi^{(b)}|\Psi^{(b)}\rangle=\int R^5dRd\Omfive |\psi_{K,\alpha}^{(b)}|^2=1,
\label{Psinorn}
\ee
with $\psi_{K,\alpha}^{(b)}(R)$ obeying a system of coupled Schr\"odinger equations,
\begin{multline}
\frac{1}{m_Q}\left(-\frac{d^2}{dR^2}+\frac{L_K(L_K+1)}{R^2}\right)\psi_{K,\alpha}^{(b)}(R)\\
+\sum_{K',\alpha'}\langle K,\alpha|V|K',\alpha'\rangle\psi_{K',\alpha'}^{(b)}(R)=E_{K,\alpha}\psi_{K,\alpha}^{(b)}(R),
\label{SchEq}
\end{multline}
where $L_K\equiv K+3/2$, and 
\be
\langle K,\alpha|V|K',\alpha'\rangle=\int d\Omfive {\cal Y}_{K,\alpha}^*(\Omfive)V(R,\Omfive){\cal Y}_{K',\alpha'}(\Omfive).
\ee
 
For the spherically symmetric ground-state solution which represents the $\Omega_{QQQ'}$ baryon studied in this work, we have, in coordinate space,
\be
\braket{{\bm\rho},{\bm\lambda}|\Omega}=\frac{1}{\pi^{3/2}R^{5/2}}\psi_0^{(b)}(R),
\ee
and, aiming at an order-of-magnitude estimate, we simplify the bound state equation (\ref{SchEq}) by neglecting all off-diagonal transitions mediated by the potential and retaining only the diagonal term, 
\be
\braket{V_S}\equiv\int\frac{dO_5}{\pi^3}V_S(\bm{\rho},\bm{\lambda}),
\label{Vsav}
\ee
so that, for the ground-state baryon, the Schr\"odinger equation (\ref{SchEq}) reduces to a one-dimensional radial equation for the wave function $\psi_0^{(b)}(R)$ formulated entirely in terms of the hyperspherical radius $R=\sqrt{\rho^2+\lambda^2}$,
\be
\frac{1}{m_Q}\left(-\frac{d^2}{dR^2}+\frac{15/4}{R^2}+\braket{V_S}\right)\psi_0^{(b)}(R)=E_\Omega\psi_0^{(b)}(R).
\label{BSeq}
\ee

From the normalization condition (\ref{Psinorn}) it is easy to find that $\psi_0^{(b)}(R)$ is normalized as a one-dimensional radial wave function, 
\be
\int_0^\infty dR \left|\psi_0^{(b)}(R)\right|^2=1. 
\ee

\subsubsection{The octet wave function}

For the continuum wave function we have
\be
\Psi_\nu=\psi_{p_\rho,p_\lambda}(\rho,\lambda){\cal Y}_\nu(\hat{\bm \rho},\hat{\bm \lambda}),
\label{Psic}
\ee
where $\nu=\{p_\rho,p_\lambda,L,L_z,l_\rho,l_\lambda\}$, with $\vep_\rho$ and $\vep_\lambda$ for the momenta conjugated to the Jacobi coordinates ${\bm\rho}$ and ${\bm\lambda}$, respectively. The angular wave function can be represented in the form
\be
{\cal Y}_\nu(\hat{\bm \rho},\hat{\bm \lambda})
=\sum_{m_\rho+m_\lambda=L_z}C_{l_\rho m_\rho j_\lambda m_\lambda}^{LL_z}Y_{l_\rho m_\rho}(\hat{\bm\rho})Y_{l_\lambda m_\lambda}(\hat{\bm\lambda}).
\ee

The radial wave function $\psi_{p_\rho,p_\lambda}(\rho,\lambda)$ is found as a continuum-spectrum solution of the equation
\be
\langle \hat{h}_{O^S}\rangle \psi_{p_\rho,p_\lambda}^S(\rho,\lambda)=E_{p_\rho,p_\lambda}\psi_{p_\rho,p_\lambda}^S(\rho,\lambda)\label{eq:HOS}
\ee
or
\be
\langle \hat{h}_{O^A}\rangle \psi_{p_\rho,p_\lambda}^A(\rho,\lambda)=E_{p_\rho,p_\lambda}\psi_{p_\rho,p_\lambda}^A(\rho,\lambda),
\label{eq:HOA}
\ee
where the superscript $A$ or $S$ is used to distinguish between the two octet representations, and
\begin{align}
\langle \hat{h}_{O^S}\rangle=\int d\hat{\bm \rho} d\hat{\bm \lambda}\, Y_{10}^*(\hat{\bm\rho})Y_{00}^*(\hat{\bm\lambda})\hat{h}_{O^S}Y_{10}(\hat{\bm\rho})Y_{00}(\hat{\bm\lambda})\nn,\\[-2mm]
\label{hOSA}\\[-2mm]
\langle \hat{h}_{O^A}\rangle=\int d\hat{\bm \rho} d\hat{\bm \lambda}\, Y_{00}^*(\hat{\bm\rho})Y_{10}^*(\hat{\bm\lambda})\hat{h}_{O^A}Y_{00}(\hat{\bm\rho})Y_{10}(\hat{\bm\lambda}).\nn
\end{align}

The radial Schr{\"o}dinger equations (\ref{eq:HOS}) and (\ref{eq:HOA}) are solved in a finite box with the length $L_{\rm box}$ in both the $\rho$ and $\lambda$ directions; see Appendix~\ref{app:SE2D} for further details of the formalism used. Thus, $\psi_{p_\rho,p_\lambda}^X(\rho,\lambda)$ ($X=A,S$) is normalized as
\be
\int \rho^2 d\rho \lambda^2 d\lambda\; \psi_{p_\rho,p_\lambda}^{X\dagger}(\rho,\lambda)\psi_{p'_\rho,p'_\lambda}^{X}(\rho,\lambda)=\delta_{p_\rho,p_\rho'}\delta_{q_\lambda,q'_\lambda}.
\ee

\begin{figure*}
\centering
\includegraphics[width=0.43\linewidth]
{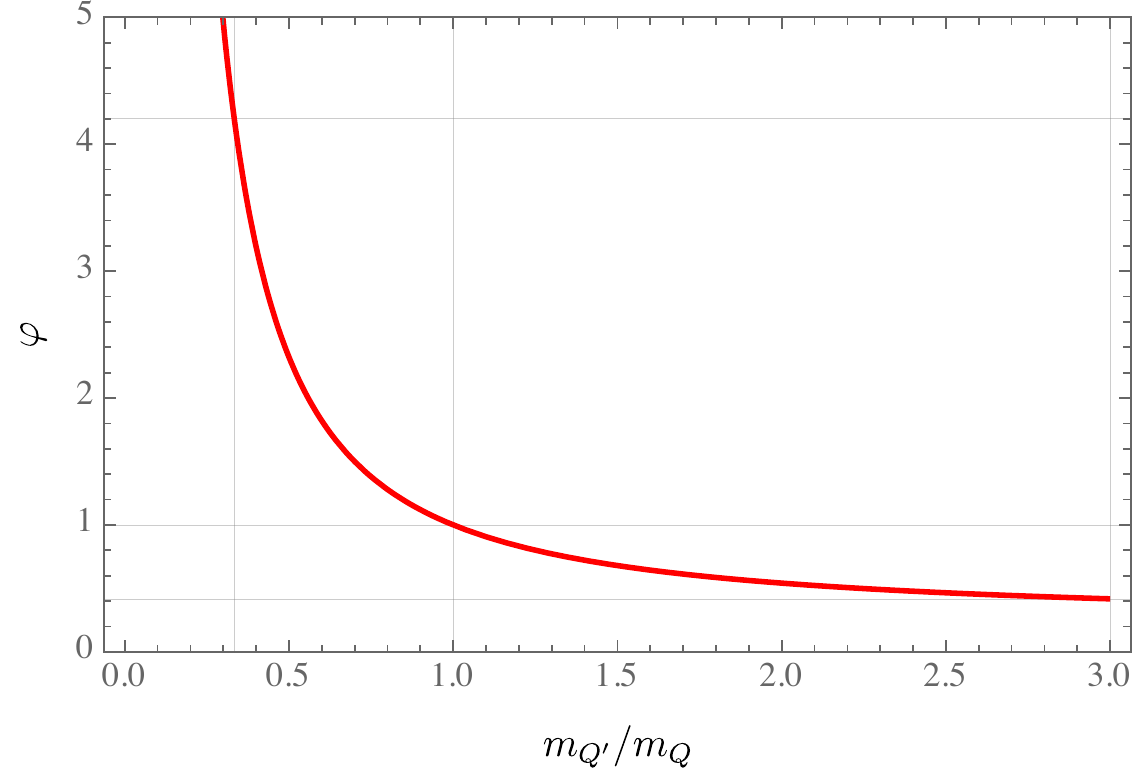}
\hspace*{0.05\linewidth}
\includegraphics[width=0.45\linewidth]
{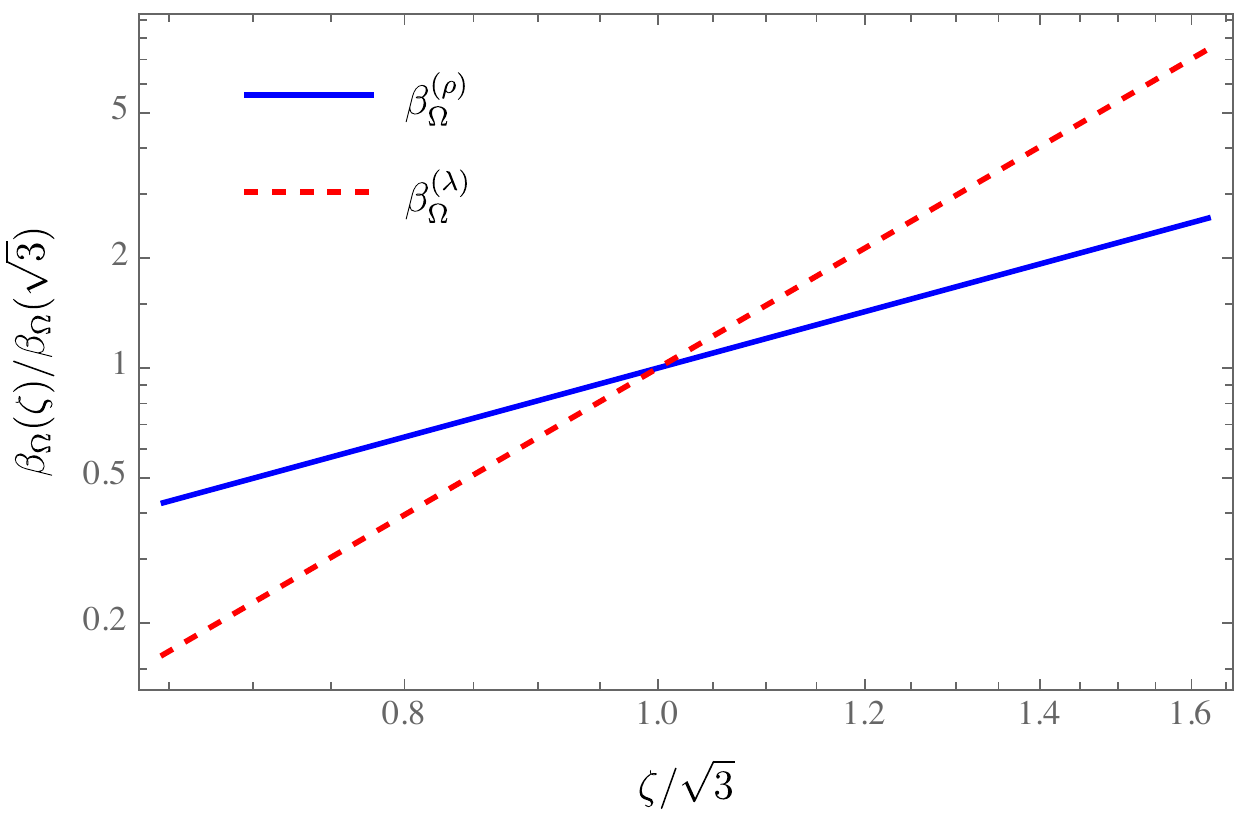}
\caption{Dependence of the correcting function $\varphi$, introduced in Eq.~(\ref{Comegadef}), on the mass ratio $m_{Q'}/m_{Q}$ (left) and the scaling behavior of $\beta_\Omega^{(\rho)}$ and $\beta_\Omega^{(\lambda)}$ from Eqs.~(\ref{sclbr}) and (\ref{sclbl}) as functions of $\zeta$ (right).}
 \label{fig:betavszeta}
\end{figure*}

\subsubsection{Numerical evaluation of $\beta_\Omega$}

Now, with both the bound-state and continuum-spectrum solutions at hand, the matrix element $\langle \Omega|\rho_z|\Psi_\nu\rangle$ can be evaluated as
\begin{align}
&I^{(\rho)}(p_\rho,p_\lambda)\equiv\langle \Omega|\rho_z|\Psi_\nu^S\rangle\nn\\
&=\int d^3\rho d^3\lambda \frac{\rho_z}{\pi^{3/2}R^{5/2}}\psi_0^{(b)}(R)\psi_{p_\rho,p_\lambda}^S(\rho,\lambda)\mathcal Y_{\nu}^S(\hat\rho,\hat\lambda)\nn\\
&=\frac{4}{\sqrt{3\pi}}\int \frac{d\rho d\lambda}{R^{5/2}}\rho^3\lambda^2\psi_0^{(b)}(R)\psi_{p_\rho,p_\lambda}^S(\rho,\lambda)
\label{Irhodef}
\end{align}
and, similarly,
\begin{align}
&I^{(\lambda)}(p_\rho,p_\lambda)\equiv\langle \Omega|\lambda_z|\Psi_\nu^A\rangle\nn\\
&=\frac{4}{\sqrt{3\pi}}\int\frac{d\rho d\lambda}{R^{5/2}}\rho^2\lambda^3\psi_0^{(b)}(R)\psi_{p_\rho,p_\lambda}^A(\rho,\lambda).
\label{Ilambdadef}
\end{align}
Further details can be found in Appendix~\ref{app:details}.

Then the two contributions to the chromopolarizability $\beta_\Omega$ take the form
\begin{align}
\beta_\Omega^{(\rho)}&=\frac1{4}\sum_{p_\rho,p_\lambda}\frac{|I^{(\rho)}(p_\rho,p_\lambda)|^2}{E_{p_\rho,p_\lambda}-E_\Omega},\\
\beta_\Omega^{(\lambda)}&=\frac{\zeta^2}{12}\sum_{p_\rho,p_\lambda} \frac{|I^{(\lambda)}(p_\rho,p_\lambda)|^2}{E_{p_\rho,p_\lambda}-E_\Omega}.
\end{align}

Similar to the case of the $\bar{Q}Q$ quarkonium, the dependence of the chromopolarizability on $m_Q$ and $\alpha_s$ can be inferred from applying the scalings in Eq.~\eqref{scales} to the definitions of the chromopolarizability of the $\Omega_{QQQ'}$ baryon in Eq.~\eqref{pol}. For the case of $m_{Q'}=m_Q$ one finds:
\begin{align}
\beta_\Omega^{(\rho)}(m_{Q'}=m_Q)&=\frac{C_{\Omega}^{(\rho)}}{m_Q^3\alpha_s^4},\label{sclbr}\\
\beta_\Omega^{(\lambda)}(m_{Q'}=m_Q)&=\frac{C_{\Omega}^{(\lambda)}}{m_Q^3\alpha_s^4}.\label{sclbl}
\end{align}
From these expressions we can expect that $\beta_\Omega^{(\rho)}$ and $\beta_\Omega^{(\lambda)}$ will be strongly dependent on the renormalization scale through the values of $m_Q$ and $\alpha_s$. Therefore, as before, we focus our attention on the values of the dimensionless coefficients $C_{\Omega}^{(\rho)}$ and $C_{\Omega}^{(\lambda)}$.

The values of $C_\Omega^{(\rho)}$ and $C_{\Omega}^{(\lambda)}$ are computed numerically by putting the system in a finite box and for $\alpha_s\in[0.3,0.5]$ and $m_Q\in[1.5,5.0]$ GeV. From these numerical computations we confirm the dependence on $\alpha_s$ and $m_Q$ in Eqs.~\eqref{sclbr} and \eqref{sclbl} and find the following values for the constants: 
\begin{align}
C_{\Omega}^{(\rho)}\approx 1.1,\,\quad C_{\Omega}^{(\lambda)}\approx1.3\,,
\end{align}
{so that}
\begin{align}
C_\Omega= C_{\Omega}^{(\rho)}+C_{\Omega}^{(\lambda)}\approx 2.4\approx 2.6\, C_\psi.
\label{ComCpsi}
\end{align}

Therefore, we conclude that for the same values of $m_Q$ and $\alpha_s$ the chromopolarizability of the $QQQ$ baryon is 2.6 times that of the $\bar{Q}Q$ meson. This results in a stronger interaction potential from the exchange of soft gluons in the double-$\Omega_{QQQ}$ system than in the double-$\bar{Q}Q$ one. 

The two main sources of the uncertainty in $C_\Omega$ can be identified. The first one, of the order of 7\%, is related to the neglect of the mixing of the color-octet representations---see Appendix~\ref{Omixing} for the details. The second source is the approximation of the fully heavy hadron by a purely Coulombic system. As discussed above, this approximation is accurate up to about 10\% for the bottom systems, but may provide a correction of the order of magnitude of the central value for the charmed systems.

In the general case of $m_{Q'}\neq m_Q$, one can parametrize the heavy quark mass dependence as
\be
\beta_\Omega=\frac{C_\Omega}{m_Q^3\alpha_s^4}\varphi(m_{Q'}/m_Q),
\label{Comegadef}
\ee
where the function $\varphi(x)$, normalized as $\varphi(1)=1$ and shown in Fig.~\ref{fig:betavszeta} (left plot), can be approximated by 
\begin{align}
\varphi(x)=\frac{C_{\Omega}^{(\rho)}}{C_\Omega}\left(\frac{\zeta(x)}{\sqrt3}\right)^{n_\rho}+\frac{C_{\Omega}^{(\lambda)}}{C_\Omega}\left(\frac{\zeta(x)}{\sqrt3}\right)^{n_\lambda},\label{eq:zetadepen}
\end{align}
with $\zeta(x)=\sqrt{1+2/x}$ (here $x=m_{Q'}/m_Q$), and $n_\rho\approx1.95$ and $n_\lambda\approx4.15$. Equation~(\ref{eq:zetadepen}) comes from the dependence of the individual contributions to $\beta_\Omega$ on the ratio $\zeta$, as shown in the right plot in Fig.~\ref{fig:betavszeta}, and is a result of a simple powerlike fit to a representative set of points calculated numerically for different values of the ratio $m_{Q'}/m_Q$. 

\begin{figure}
\centering
\includegraphics[width=0.98\linewidth]{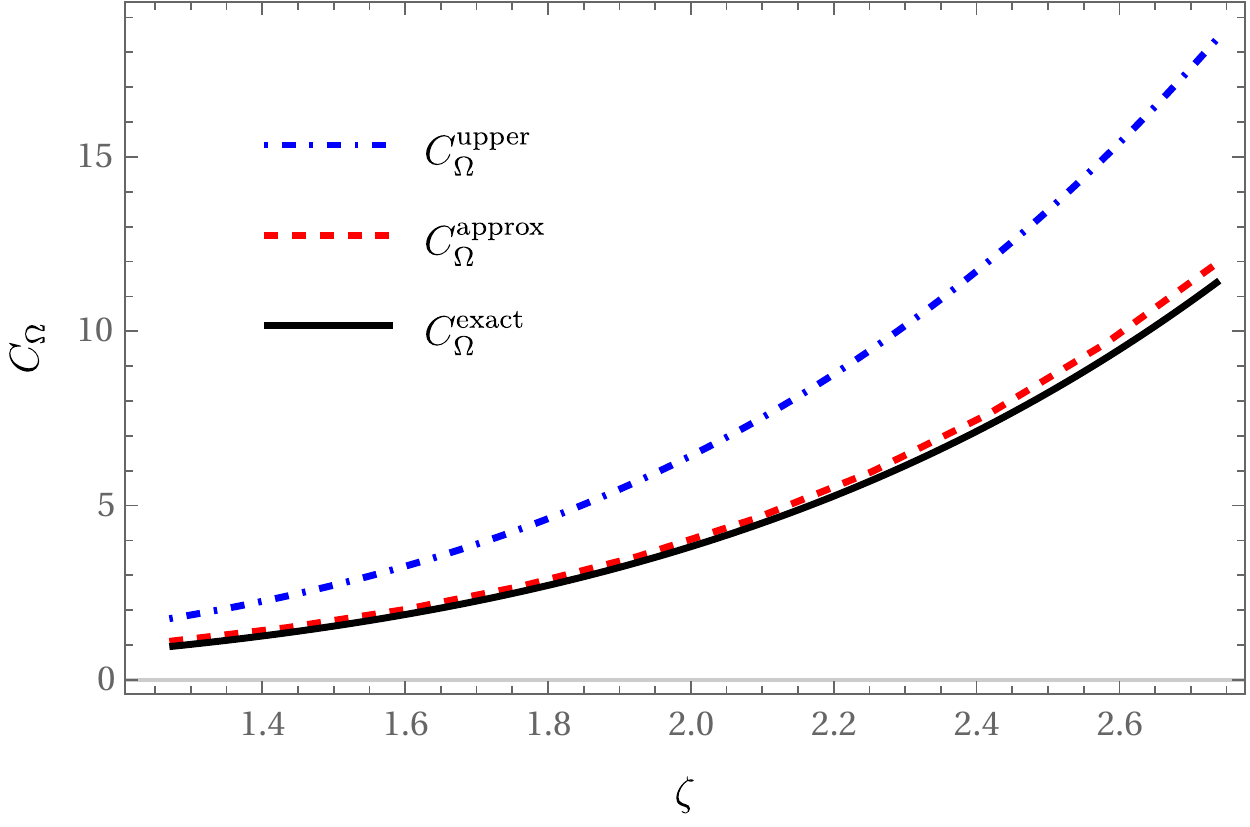}
\caption{The constant $C_\Omega$ as a function of $\zeta$: the upper bound as given in Eq.~(\ref{COupper}) (the blue dotted-dashed curve), the exact result quoted in Eq.~(\ref{Comegadef}) (the black solid curve), and the approximate value provided by Eq.~(\ref{pol3}) (the red dashed curve).}
\label{fig:COmega} 
\end{figure}

The upper bound on the chromopolarizability $\beta_\Omega$ can be estimated by employing the same approach as in the case of the quarkonium to arrive at the result
\be
\beta_\Omega^{\rm upper}=\frac{\braket{R^2}(\zeta^2+3)}{72|E_\Omega|},
\label{COupper}
\ee
which is shown as the dotted-dashed blue curve in Fig.~\ref{fig:COmega} while the exact result (\ref{Comegadef}) is given by the black solid curve. 

It is also instructive to derive approximate
formulas for the chromopolarizabilities $\beta_\Omega^{(\rho)}$ and $\beta_\Omega^{(\lambda)}$ of a fully heavy baryon employing the same technique as in the case of the quarkonium discussed above. To this end, we consider approximate expressions following from Eq.~(\ref{pol}),
\be
\begin{split}
\beta_\Omega^{(\rho)}&\approx\frac{1}{12}\frac{\braket{\rho^2}}{\braket{\hat{T}({\bm p}_\rho,{\bm p}_\lambda)}+\braket{V_{O^S}({\bm\rho},{\bm\lambda})}-E_\Omega},\\
\beta_\Omega^{(\lambda)}&\approx
\frac{\zeta^2}{36}\frac{\braket{\lambda^2}}{\braket{\hat{T}({\bm p}_\rho,{\bm p}_\lambda)}+\braket{V_{O^A}({\bm\rho},{\bm\lambda})}-E_\Omega},
\label{pol3}
\end{split}
\ee
where $\braket{\ldots}$ stands for the averaging over the ground state $\ket{\Omega}$. Given the empirical relation derived for the quarkonium, it is natural to expect that each degree of freedom (in $\rho$ and in $\lambda$) will contribute an equal amount $|E_\Omega|/2$ to the averaged kinetic energy, so that $\braket{\hat{T}({\bm p}_\rho,{\bm p}_\lambda)}\approx|E_\Omega|$ in total. Indeed, with this substitution, the approximate result for the coefficient $C_\Omega^{\rm approx}$ which follows straightforwardly from Eq.~(\ref{pol3}) agrees very well with the exact result from Eq.~(\ref{Comegadef})---see the red dashed and black solid curves in Fig.~\ref{fig:COmega}, respectively.

Similar to the case of the heavy quarkonium studied in Sec.~\ref{sec:betaofbb}, the dependence of the chromopolarizability $\beta_\Omega$ from Eq.~(\ref{Comegadef}) on $m_Q$ and $\alpha_s$ is quite strong. As mentioned above, the estimate from Eq.~(\ref{ComCpsi}) is obtained under the assumption that the same quark mass and strong coupling constant are used in both cases, for $\beta_\psi$ and $\beta_\Omega$. Numerical values of the fully heavy baryon chromopolarizabilities for the values of $m_c$, $m_b$, and $\alpha_s$ quoted in Eqs.~(\ref{eq:masses}) and (\ref{eq:alphas}) and discussed in Sec.~\ref{sec:numpsi}, are listed in Table~\ref{tab:betas}. 

\begin{table}[t]
\caption{The mean radii and chromopolarizabilities of the fully heavy $QQQ'$ baryons with $Q,Q'=c,b$ evaluated in this work. See Sec.~\ref{sec:numpsi} for the discussion of the uncertainties.}
\label{tab:betas}
\begin{ruledtabular}
\renewcommand\arraystretch{1.7}
\begin{tabular}{lcccc}
State & $ccc$ & $ccb$ & $cbb$ & $bbb$ \\ 
\hline
$\langle R\rangle$ [fm] & $1.66_{-0.46}^{+0.26}$ & $1.44_{-0.40}^{+0.23}$ &$0.65_{-0.18}^{+0.10}$ & $0.51_{-0.14}^{+0.08}$ \\
$\beta_\Omega$ [GeV$^{-3}$] & $49_{-35}^{+38}$ & $19_{-14}^{+15}$ &$6.7_{-4.8}^{+5.3}$ & $1.4_{-1.0}^{+1.1}$ 
\end{tabular}
\end{ruledtabular}
\end{table}

\section{Summary and discussions}
\label{sec:concl}

In this paper we evaluate the chromopolarizability of a fully heavy ground-state baryon $\Omega_{QQQ}$ and find it to be $2.6$ of that for the heavy meson $\bar{Q}Q$ composed of the quark and its antiquark of the same flavor. To estimate the uncertainty of this result we notice that the approximation of a fully heavy hadron by a purely color-Coulombic system employed in this study predictably works well for the ground-state heavy quarkonium and baryon composed of the bottom quarks, that is, for $\bar{b}b$ and $bbb$ systems. Indeed, the corresponding mean radii quoted in Tables~\ref{tab:betaspsi} and \ref{tab:betas} appear fairly small compared with the scale $r_{\rm np}\sim 1/\sqrt{\sigma}\sim 1/\Lambda_{\rm QCD}\simeq 0.7$~fm which roughly quantifies the relevance of the nonperturbative interaction in the hadron. Then, for such systems, we sum in quadrature the uncertainties which come from the nonperturbative dynamics (about 10\%) and from neglecting the off-diagonal chromopolarizabilites (about 7\%) to obtain
\be
\beta_{\Omega_{bbb}}/\beta_{\bar{b}b}\approx 2.6\pm 0.3.
\ee 

In the meantime, corrections from the nonperturbative interaction may appear at the level of the leading order for the $\bar{c}c$ charmonium and fully-charmed baryon $\Omega_{ccc}$. Therefore, for such systems, the ratio above should be regarded as an order-of-magnitude estimate. 

We further extend our analysis to fully heavy baryons containing different flavors of heavy quarks, $Q$ and $Q'$, and discuss the dependence of the chromopolarizability of the baryon $QQQ'$ on the mass ratio $m_{Q'}/m_Q$. We notice that the system $cbb$ appears to be rather compact, with the mean size of the same order as that of the $bbb$ one, so corrections from the nonperturbative dynamics for such system are expected to be at the same level of 10\% or so. In the meantime, the $ccb$ system is rather large, like the $ccc$ one, so we pretend to provide only an order-of-magnitude estimate for it. Nevertheless, we still expect our results for the chromopolarizabilities of such baryons to lie in the right ballpark, especially given the large estimated uncertainty in $\beta_\Omega$ related to those in the quark mass and strong coupling constant determination. 

Our findings imply that the interaction from the exchange of soft gluons in the double-$\Omega_{QQQ}$ system appears to be considerably stronger than that in the double-heavy-quarkonium pair and, as such, provides an argument in favor of the existence of near-threshold (bound or virtual) states in such a dibaryon system. It is interesting to notice the fact that lattice calculations indeed report a possible existence of bound states in both double-$\Omega_{ccc}$ \cite{Lyu:2021qsh} and double-$\Omega_{bbb}$ \cite{Mathur:2022nez} systems, with the binding energies of the order of several MeV, in the former case (when the electric Coulomb repulsion is neglected), and several dozen MeV, in the latter.

It has to be noticed that the current situation with the experimental studies of the properties of fully heavy baryons does not look very promising even in the charm sector at the present stage; the existing experimental data are limited to some candidates for the singly charmed baryons~\cite{ParticleDataGroup:2022pth} and only one doubly charmed baryon reported by the LHCb Collaboration~\cite{LHCb:2017iph}. However, on the one hand, the integrated luminosity of LHCb at the end of the planned Run 5 will reach 300~fb$^{-1}$~\cite{Cerri:2018ypt}, two orders of magnitude higher than 1.7~fb$^{-1}$ of Run 1 from which the observation of the $\Xi_{cc}^{++}$ was made~\cite{LHCb:2017iph}. According to the calculations in Ref.~\cite{Chen:2011mb}, it is promising to observe the $\Omega_{ccc}$ and $\Omega_{ccb}$ in the data of future runs of the Large Hadron Collider. On the other hand, there is some progress in lattice studies of the mass and electromagnetic form factor of the fully-charmed baryons 
\cite{Padmanath:2013zfa,Can:2015exa} as well as double-$\Omega_{ccc}$ \cite{Lyu:2021qsh} and double-$\Omega_{bbb}$ \cite{Mathur:2022nez} systems. As favored by our findings reported in this work, further experimental and lattice searches for near-threshold hadronic molecules of fully heavy hadrons would look like an appealing and promising task and are very likely to result in discoveries of new two-hadron resonances.

\section*{Acknowledgments}
The authors would like to thank A.Vairo for reading the manuscript and valuable comments. X.-K.D is grateful to the helpful discussions with Wei Hao, Teng Ji, and Qiang Zhao. This work was supported in part by the Chinese Academy of Sciences under Grant No.~XDB34030000; by the National Natural Science Foundation of China (NSFC) under Grants No.~12125507, No. 11835015, No.~12047503, and No.~11961141012; and by the NSFC and the Deutsche Forschungsgemeinschaft (DFG) through the funds provided to the Sino-German Collaborative Research Center TRR110 ``Symmetries and the Emergence of Structure in QCD'' (NSFC Grant No.~12070131001, DFG Project-ID~196253076). A.N. is supported by the Slovenian Research Agency (research core funding No. P1-0035). J.T.C. acknowledges financial support by National Science Foundation (No. PHY-2013184). 

\appendix

\section{Generalized Jacobi coordinates}
\label{app:Jacobi}

In this appendix we provide a generalized form of the Jacobi coordinates in a 3-body system. We follow Ref.~\cite{FabredelaRipelle:1991aa}.
In particular, the center-of-mass coordinate $\veX$ and the two relative coordinates, $\velambda$ and $\verho$, can be introduced as
\begin{align}
\vex_1&=\veX+\sqrt{\frac{\mu m_3}{M(m_1+m_2)}}\velambda+\sqrt{\frac{\mu m_2}{m_1(m_1+m_2)}}\verho,\nn\\
\vex_2&=\veX+\sqrt{\frac{\mu m_3}{M(m_1+m_2)}}\velambda-\sqrt{\frac{\mu m_1}{m_2(m_1+m_2)}}\verho,\nn\\
\vex_3&=\veX-\sqrt{\frac{\mu(m_1+m_2)}{Mm_3}}\velambda,\label{x1x2x3}
\end{align}
where $\vex_i$ ($i=1,2,3$) are the individual coordinates of the particles, $M=m_1+m_2+m_3$, and $\mu$ is an arbitrary parameter with the dimension of mass. The inverse of Eq.~(\ref{x1x2x3}) reads
\begin{align}
\veX&=\frac{1}{M}\sum_{i=1}^3m_i\vex_i,\nn\\
\velambda&=\sqrt{\frac{(m_1+m_2)m_3}{\mu M}}\left(\frac{m_1\vex_1+m_2\vex_2}{m_1+m_2}-\vex_3\right),\label{Xrl}\\
\verho&=\sqrt{\frac{m_1m_2}{\mu(m_1+m_2)}}(\vex_1-\vex_2),\nn
\end{align}
so that $\verho$ can be regarded as the relative coordinate of the particles 1 and 2 while $\velambda$ describes the separation between particle 3 and the center-of-mass of particles 1 and 2.

The nonrelativistic kinetic term in the Lagrangian turns to be
\be
\frac12\sum_{i=1}^3m_i\dot{\vex}_i^2=\frac12M\dot{\bm{X}}^2+\frac12\mu\left(\dot{\bm{\lambda}}^2+\dot{\bm{\rho}}^2\right),
\label{lagkin}
\ee
where the arbitrary parameter $\mu$ plays the role of the mass for both motions in ${\bm\rho}$ and ${\bm\lambda}$, which makes the coordinate transformation (\ref{x1x2x3}) particularly convenient in practical calculations. The physical observables do not depend on a particular choice of $\mu$, so for convenience, in case of $m_1=m_2=m_Q$ and $m_3=m_{Q'}$, we set $\mu=m_Q/2$, so that Eqs.~(\ref{Xrl}) and (\ref{lagkin}) turn to Eqs.~(\ref{eq:Jacobi}) and (\ref{lagkin0}), respectively. As an additional check we have verified that the results reported in this paper do not depend on a particular choice of $\mu$, as required.

\section{Details of the $\beta_\Omega$ calculations}
\label{app:details}

In this appendix we collect some formulas used in Sec.~\ref{sec:betaomega} to evaluate the chromopolarizability of the fully heavy baryon $\beta_\Omega$. 

We start from the averaged value of the interaction potential in the color-singlet representation defined in Eq.~(\ref{Vsav}) to find that
\begin{align}
\braket{V_S}&=-\frac{2\alpha_s}{3}\int \frac{d\Omfive}{\pi^3} \left(\frac{1}{|\bm{\rho}|}+\frac{2}{|\zeta\bm{\lambda}+\bm{\rho}|}+\frac{2}{|\zeta\bm{\lambda}-\bm{\rho}|}\right)\nn\\
&=
-\frac{32\alpha_s}{9\pi R}\left(1+\frac{4}{\sqrt{1+\zeta^2}}\right),
\end{align}
where an easily verifiable master formula,
$$
\left\langle |a{\bm\rho}+b{\bm\lambda}|^{-1}\right\rangle=\int\frac{d\Omfive}{\pi^3}\frac1{|a{\bm\rho}+b{\bm\lambda}|}=\frac{16}{3\pi R\sqrt{a^2+b^2}},
$$
was used, which is valid for arbitrary numerical coefficients $a$ and $b$. The averaged Hamiltonians for the octet representations as defined in Eq.~(\ref{hOSA}) are evaluated as
\begin{align}
\langle \hat{h}_{O^S}\rangle&=\int d\hat{\bm \rho} d\hat{\bm \lambda} Y_{10}^*(\hat{\bm\rho})Y_{00}^*(\hat{\bm\lambda})\hat{h}_{O^S}Y_{10}(\hat{\bm\rho})Y_{00}(\hat{\bm\lambda})\nn\\
&=\int d\hat{\bm \rho} |Y_{10}(\hat{\bm\rho})|^2 \int \frac{d\hat{\bm \lambda}}{4\pi}\hat{h}_{O^S}\nn\\
&=\int\frac{d{\hat{\bm\rho\bm\lambda}}}{4\pi}\hat{h}_{O^S}[l_\rho=1,l_\lambda=0]\nn\\
&=\frac{1}{m_Q}\left(-\frac{1}{\rho}\frac{\partial^2}{\partial\rho^2}\rho-\frac{1}{\lambda}\frac{\partial^2}{\partial\lambda^2}\lambda
+\frac{2}{\rho^2}\right)\nn\\
&\ \ \ \ 
+\frac{\alpha_s}{3}\left(\frac{1}{\rho}-\frac{5}{\mbox{max}(\rho,\zeta\lambda)}\right)
\label{hSOav}
\end{align}
and, similarly,
\begin{align}
\langle \hat{h}_{O^A}\rangle&=\int d\hat{\bm \rho} d\hat{\bm \lambda} Y_{00}^*(\hat{\bm\rho})Y_{10}^*(\hat{\bm\lambda})\hat{h}_{O^A}Y_{00}(\hat{\bm\rho})Y_{10}(\hat{\bm\lambda})\nn\\
&=\frac{1}{m_Q}\left(-\frac{1}{\rho}\frac{\partial^2}{\partial\rho^2}\rho-\frac{1}{\lambda}\frac{\partial^2}{\partial\lambda^2}\lambda
+\frac{2}{\lambda^2}\right)\nn\\
&\ \ \ \ 
-\frac{2\alpha_s}{3}\left(\frac{1}{\rho}-\frac{1}{2\rm{max}(\rho,\zeta\lambda)}\right),
\label{hSAav}
\end{align}
where the following master formula was used:
$$
\frac{1}{4\pi}\int\frac{d{\hat{\bm v\bm w}}}{|{\bm v}+{\bm w}|}=\frac12\left(\left|\frac1v+\frac1w\right|-\left|\frac1v-\frac1w\right|\right)=\frac{1}{\mbox{max}(v,w)}.
$$

Finally, the evaluation of the angular part of the integral in Eq.~(\ref{Irhodef}) is done as
\begin{widetext}
\bea
&&\sum_{m_\rho,m_\lambda}C_{l_\rho m_\rho j_\lambda m_\lambda}^{LL_z} \int d\hat{\bm \rho} d\hat{\bm \lambda}\; \hat{\rho}_z Y_{l_\rho m_\rho}(\hat{\bm\rho})Y_{l_\lambda m_\lambda}(\hat{\bm\lambda})
=4\pi\sum_{m_\rho,m_\lambda}C_{l_\rho m_\rho j_\lambda m_\lambda}^{LL_z} \int d\hat{\bm \rho} d\hat{\bm \lambda}\; 
Y_{00}(\hat{\bm\rho})Y_{00}(\hat{\bm\lambda})\hat{\rho}_z Y_{l_\rho m_\rho}(\hat{\bm\rho})Y_{l_\lambda m_\lambda}(\hat{\bm\lambda})
\nn\\
&=&4\pi\sum_{m_\rho,m_\lambda}C_{l_\rho m_\rho j_\lambda m_\lambda}^{LL_z} \langle L=0,L_z=0,l_\rho=0,l_\lambda=0|\hat{\rho}_z|L,L_z,l_\rho,l_\lambda\rangle
=\delta_{L1}\delta_{L_z0}\delta_{l_\rho1}\delta_{l_\lambda0}\langle l_\rho=0|\hat{\rho}_z|l_\rho=1\rangle\\
&=&\frac{4\pi}{\sqrt{3}}\delta_{L1}\delta_{L_z0}\delta_{l_\rho1}\delta_{l_\lambda0},\nn
\eea
\end{widetext}
where it was used that, for an arbitrary 3-vector ${\bm v}$,
\be
\langle l-1,m|v_z|l,m\rangle=\sqrt{\frac{l^2-m^2}{l(2l-1)(2l+1)}}\langle l-1||v_z||l\rangle,
\ee
and the reduced matrix element for $v_z=n_z$ reads
\be
\langle l-1||n_z||l\rangle=\sqrt{l}.
\ee
A similar calculation was performed for the integral defined in Eq.~(\ref{Ilambdadef}).

\section{Evaluation of the $QQQ'$ continuum spectrum in a 2D well}\label{app:SE2D}

In this appendix we provide some technical details related to building the continuum spectrum of the Hamiltonians $\hat{h}_{O^S}$ and $\hat{h}_{O^A}$ defined in Eqs.~(\ref{eq:HOS}) and (\ref{eq:HOA}).

Employing the same technique as was previously used to build the continuum spectrum of the Hamiltonian $\hat{h}_O$ in Sec.~\ref{sec:betapsi}, we embed the 
studied three-quark system in a finite rectangle box of the size $L_{\rm box}$ in each spatial direction, so that for $0<\rho,\lambda<L_{\rm box}$ the potentials (\ref{hOS}) and (\ref{hOA}) averaged as defined in Eq.~(\ref{hOSA}) [see also Eqs.~(\ref{hSOav}) and (\ref{hSAav})] take the form
\be
V_{O^S}(\rho,\lambda)=\frac{2}{m_Q \rho^2}
+\frac{\alpha_s}{3}\left(\frac{1}{\rho}-\frac{5}{\mbox{max}(\rho,\zeta\lambda)}\right)
\ee
and
\be
V_{O^A}(\rho,\lambda)=\frac{2}{m_Q \lambda^2}
-\frac{2\alpha_s}{3}\left(\frac{1}{\rho}-\frac{1}{2\mbox{max}(\rho,\zeta\lambda)}\right),
\ee
respectively.

It is convenient then to define a length scale 
$L_0=0.197$~fm ($1/L_0=1$ GeV) and proceed to the dimensionless Jacobi coordinates, $\tilde{\rho}=\rho/L_0$ $\tilde{\lambda}=\lambda/L_0$, as well as other quantities relevant for the calculation,
\be
\tilde{L}_{\rm box}=L_{\rm box}/L_0,\quad \tilde{m}_{Q^{(\prime)}}=m_{Q^{(\prime)}}L_0,\quad \tilde{E}=EL_0.
\ee
It also proves convenient to define a radial wave function $\chi=\rho\lambda\psi$ and its dimensionless counterpart $\tilde{\chi}$ which is
normalized as
\begin{align}
\int_0^{\tilde{L}_{\rm box}}d\tilde\rho \int_0^{\tilde{L}_{\rm box}}d\tilde\lambda\,|\tilde\chi(\tilde\rho,\tilde\lambda)|^2=1
\end{align}
and obeys the radial Schr{\"o}dinger equation 
\begin{align}
\frac1{\tilde{m}_Q}\left( -\frac{\partial^2}{\partial \tilde{\rho}^2}- \frac{\partial^2}{\partial \tilde{\lambda}^2}+\tilde V(\tilde \rho,\tilde\lambda)\right)\tilde\chi(\tilde \rho,\tilde\lambda)=\tilde E\tilde\chi(\tilde \rho,\tilde\lambda),\label{eq:pdeu}
\end{align}
with
\begin{align}
\tilde V(\tilde \rho,\tilde\lambda)=\frac{2}{\tilde \rho^2}+\frac13\alpha_s\tilde{m}_Q \left(\frac1{\tilde \rho}-\frac{5}{\mbox{max}(\tilde\rho,\zeta\tilde\lambda)}\right)
\end{align}
or
\begin{align}
\tilde V(\tilde \rho,\tilde\lambda)=\frac{2}{\tilde \lambda^2}-\frac23\alpha_s\tilde{m}_Q\left(\frac1{\tilde \rho}-\frac{1}{2\mbox{max}(\tilde\rho,\zeta\tilde\lambda)}\right),
\end{align}
depending on which contribution to $\beta_\Omega$ is considered. 

The boundary conditions imposed on the wave function $\tilde{\chi}$ read
\begin{align}
\tilde\chi(0,\tilde\lambda)
=\tilde\chi(\tilde{L}_{\rm box},\tilde\lambda)
=\tilde\chi(\tilde\rho,0)
=\tilde\chi(\tilde\rho,\tilde{L}_{\rm box})
=0.
\end{align}
Equation (\ref{eq:pdeu}) is then solved numerically using the spectrum method of Ref.~\cite{Andrade:2017jmt}.

\section{Diagonalization of the octet fields}
\label{Omixing}

\begin{figure*}
\centering
\includegraphics[width=0.47\linewidth]{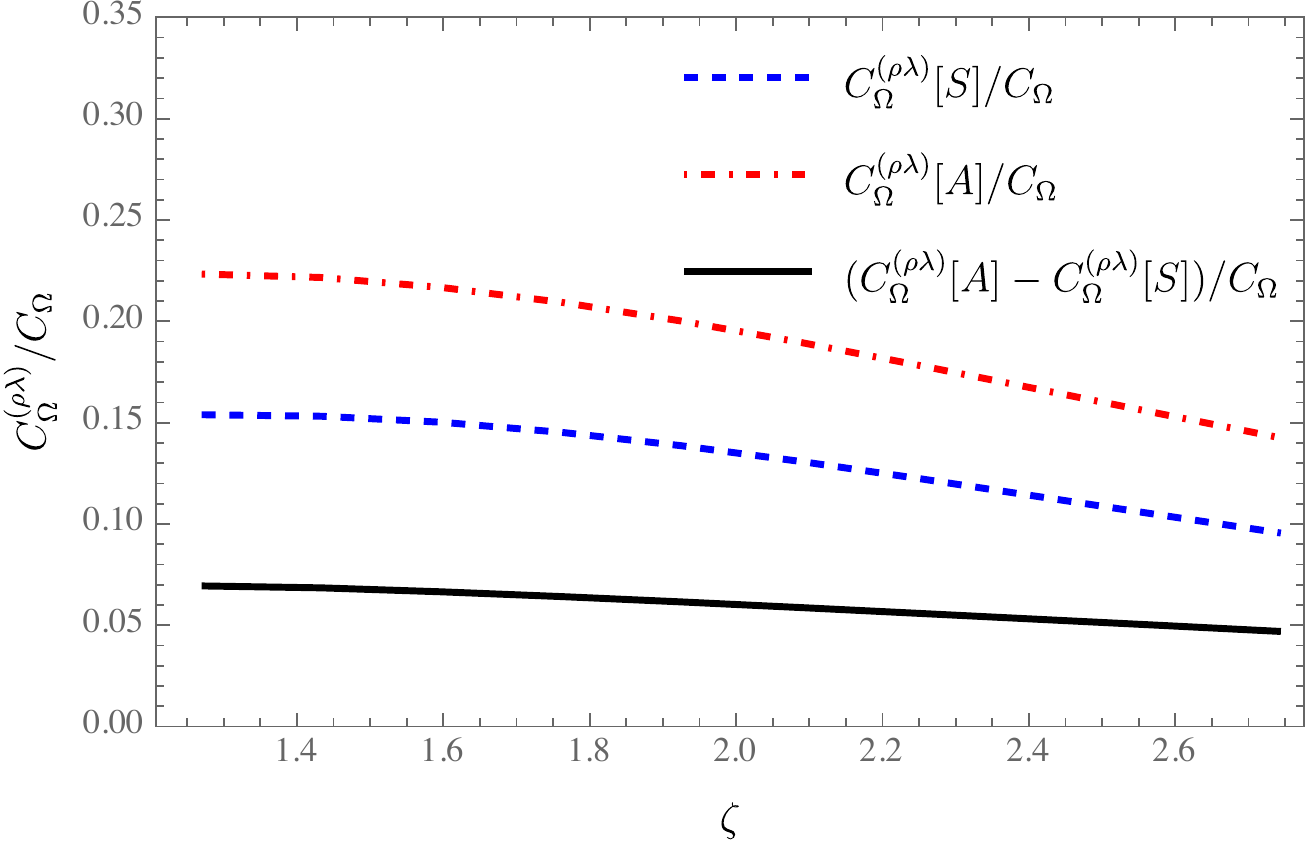}
\hspace*{0.03\linewidth}
\includegraphics[width=0.455\linewidth]{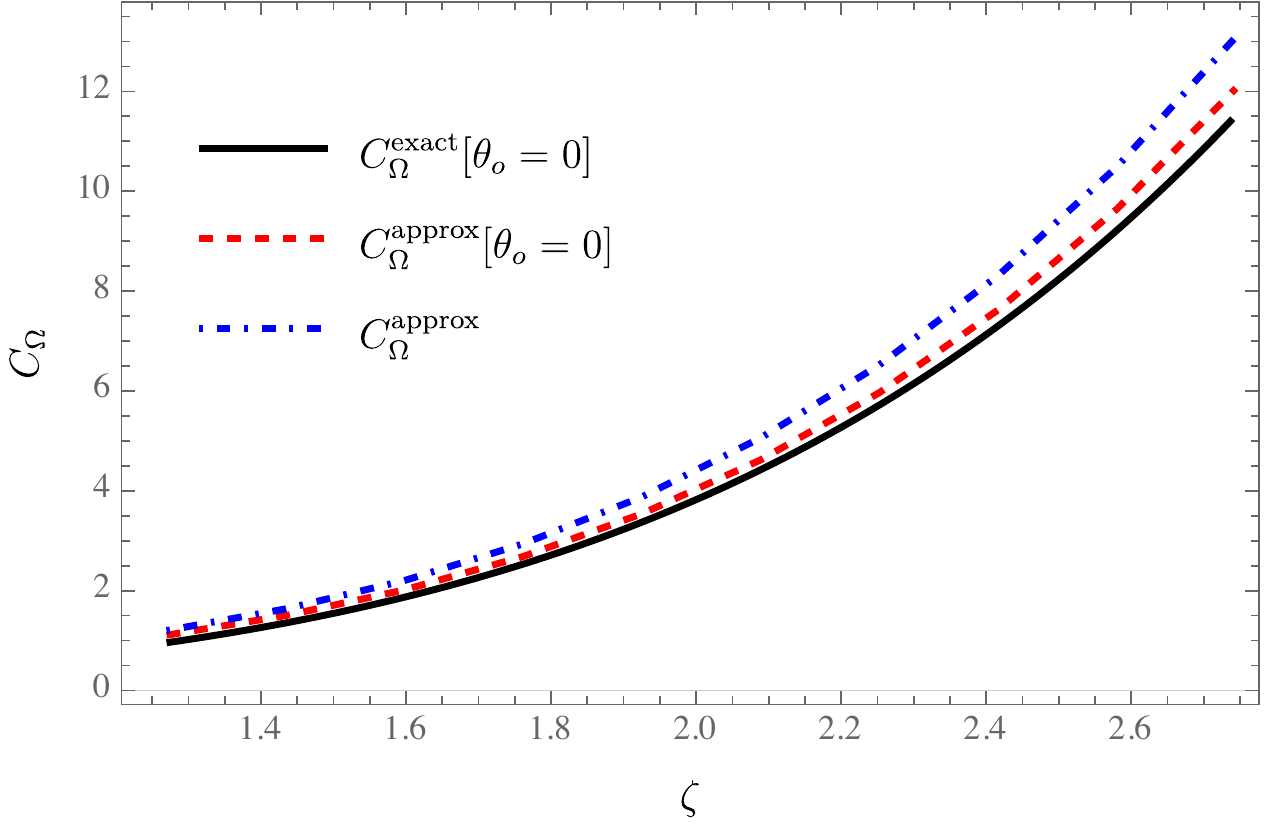}
\caption{Left plot: the ratio of the mixed coefficient $C_\Omega^{(\rho\lambda)}$ to the full coefficient $C_\Omega$ (the black solid curve) and the individual contributions to this ratio. Right plot: the approximate coefficient $C_\Omega$ evaluated for the zero (the red dashed curve) and nonzero (the blue dot curve) mixing compared with the exact result obtained for zero mixing---see Eq.~(\ref{Comegadef}) (the black solid curve).}
\label{fig:Crl} 
\end{figure*}

In this appendix we discuss the effect of the octet fields mixing for the fully heavy baryon $\Omega$. The effective Lagrangian at leading order reads
\begin{align}
{\cal L}^{(0)}_{\rm pNRQCD}=&\int d^3\rho d^3\lambda\left\{S^\dag[i\partial_0-V^s]S+\Delta^\dag[i\partial_0-V^s]\Delta\right.\nonumber\\
&\left.+\bm{O}^\dag[iD_0-\bm{V}^o]\bm{O}\right\}\,\label{LOO},
\end{align}
where $\bm{O}=(O^{A}, O^{S})$ and 
\begin{align}
\bm{V}^o=\left(\begin{array}{cc} V_{O^A} & V_{O^{AS}} \\ V_{O^{AS}} & V_{O^S}\end{array}\right).
\end{align}
The diagonal octet potentials $V_{O^A}$ and $V_{O^A}$ are quoted in Eqs.~(\ref{hOS}) and (\ref{hOA}), respectively, and the off-diagonal mixing term reads \cite{Brambilla:2013vx}
\begin{align}
V_{O^{AS}}&=-\frac{\sqrt{3}\alpha_s}{2}\left(\frac{1}{|\bm{\rho}+\zeta\bm{\lambda}|}-\frac{1}{|\bm{\rho}-\zeta\bm{\lambda}|}\right).
\label{eq:VAS}
\end{align}

To deal with the octets mixing term one can follow two equivalent approaches. One possibility is to resum the mixing potential insertions in the octet propagators \cite{Brambilla:2013vx}. The other possibility, which we employ below, is to diagonalize the octet potential matrix. Thus we define the rotation matrix,
\begin{align}
R=\left(\begin{array}{cc} \cos\theta_o & -\sin\theta_o \\ \sin\theta_o & \cos\theta_o \end{array}\right),
\end{align}
which diagonalizes the potential matrix,
\begin{align}
R^T\bm{V}R={\rm diag}(V_+,V_-).
\end{align}
Then it is easy to find that
\begin{eqnarray}
\sin2\theta_o&\ds=\frac{2V_{O^{AS}}}{\sqrt{(V_{O^A}-V_{O^S})^2+4V_{O^{AS}}^2}},\nn\\[-2mm]
\label{sc2t}\\[-2mm]
\cos2\theta_o&\ds =\frac{V_{O^A}-V_{O^S}}{\sqrt{(V_{O^A}-V_{O^S})^2+4V_{O^{AS}}^2}}\nn,
\end{eqnarray}
and
\begin{align}
V_\pm=&\frac{1}{2}\left(V_{O^A}+V_{O^S}\pm\sqrt{(V_{O^A}-V_{O^S})^2+4V_{O^{AS}}^2}\right).
\end{align}

Now we can express the interaction Lagrangian in Eq.~\eqref{dlg} in terms of the rotated octet fields by using $\bm{O}=R\widetilde{\bm{O}}$,
\begin{align}
&{\cal L}_{\rm pNRQCD}^{\mbox{\scriptsize $S$-$O$}}=\int d^3\rho d^3\lambda\left\{\frac{1}{2\sqrt{2}}\left[ S^\dag\bm{\rho}\cdot g\bm{E}^a\left(\cos\theta_o \widetilde{O}^{S a}\right.\right.\right.\nonumber\\
&\left.\left.\left.+\sin\theta_o \widetilde{O}^{A a }\right)+H.c.\right]-\frac{\zeta}{2\sqrt{6}}\left[S^\dag\bm{\lambda}\cdot g\bm{E}^a\left(\cos\theta_o \widetilde{O}^{A a}\right.\right.\right.\nonumber\\
&\left.\left.\left.-\sin\theta_o \widetilde{O}^{S a}\right)+H.c.\right]\right\}.
\end{align}

Then for the chromopolarizability we find
\begin{align}
\beta_\Omega=\tilde{\beta}_\Omega^{(\rho)}+\tilde{\beta}_\Omega^{(\lambda)}+\tilde{\beta}_\Omega^{(\rho \lambda)},
\end{align}
where
\begin{align}
\tilde{\beta}_\Omega^{(\rho)}=&\frac{1}{12}\bra{\Omega}\bm{\rho}\left(\sin\theta_o\frac{1}{\hat{h}_{\widetilde{O}_A}-E_\Omega}\sin\theta_o\right.\nonumber\\
&\left.+\cos\theta_o\frac{1}{\hat{h}_{\widetilde{O}_S}-E_\Omega}\cos\theta_o\right)\bm{\rho}\ket{\Omega},
\label{br}\\
\tilde{\beta}_\Omega^{(\lambda)}=&
\frac{\zeta^2}{36}\langle \Omega|\bm{\lambda}\left(\cos\theta_o\frac{1}{\hat{h}_{\widetilde{O}_A}-E_\Omega}\cos\theta_o\right.\nonumber\\
&\left.+\sin\theta_o\frac{1}{\hat{h}_{\widetilde{O}_S}-E_\Omega}\sin\theta_o\right)\bm{\lambda}\ket{\Omega},\label{bl}\\
\tilde{\beta}_\Omega^{(\rho \lambda)}=&
-\frac{\zeta}{6\sqrt{3}}\langle \Omega|\bm{\rho}\left(\sin\theta_o\frac{1}{\hat{h}_{\widetilde{O}_A}-E_\Omega}\cos\theta_o\right.\nonumber\\
&\left.-\cos\theta_o\frac{1}{\hat{h}_{\widetilde{O}_S}-E_\Omega}\sin\theta_o\right)\bm{\lambda}\ket{\Omega},
\end{align}
and the Hamiltonians take the form
\begin{align}
\hat{h}_{\widetilde{O}_A}&=\hat{T}(\vep_\rho,\vep_\lambda)+V_+({\bm\rho},{\bm\lambda}),\nn\\[-2mm]
\label{tildeH}\\[-2mm]
\hat{h}_{\widetilde{O}_S}&=\hat{T}(\vep_\rho,\vep_\lambda)+V_-({\bm\rho},{\bm\lambda}).\nn
\end{align}

Consider first the mixed term, 
\be
\tilde{\beta}_\Omega^{(\rho \lambda)}=\tilde{\beta}_\Omega^{(\rho \lambda)}[A]-\tilde{\beta}_\Omega^{(\rho \lambda)}[S],
\label{bAS}
\ee
where 
\be
\tilde{\beta}_\Omega^{(\rho \lambda)}[X]=
- \frac{\zeta}{12\sqrt{3}}\bra{\Omega}\frac{{\bm\rho}\cdot {\bm \lambda}\sin2\theta_o}{\hat{h}_{\widetilde{O}_X}-E_\Omega}\ket{\Omega},
\ee
with $X=A,S$. Then, employing the same approximate approach as in Eq.~(\ref{pol3}), we can write
\begin{eqnarray}
\tilde{\beta}_\Omega^{(\rho\lambda)}[S]\approx-\frac{\zeta}{12\sqrt{3}}\frac{\braket{\bm{\rho}\cdot {\bm\lambda}\sin2\theta_o}}{\braket{\hat{T}({\bm p}_\rho,{\bm p}_\lambda)}+\braket{V_-({\bm\rho},{\bm\lambda})}-E_\Omega},\nn\\[0mm]
\label{pol4}\\[-2mm]
\tilde{\beta}_\Omega^{(\rho\lambda)}[A]\approx
-\frac{\zeta}{12\sqrt{3}}\frac{\braket{\bm{\rho}\cdot \bm{\lambda}\sin2\theta_o}}{\braket{\hat{T}({\bm p}_\rho,{\bm p}_\lambda)}+\braket{V_+({\bm\rho},{\bm\lambda})}-E_\Omega},\nn
\end{eqnarray}
where, as before, $\braket{\ldots}$ stands for the averaging over the ground state $\ket{\Omega}$ and $\braket{\hat{T}({\bm p}_\rho,{\bm p}_\lambda)}=|E_\Omega|$. The corresponding coefficients $C_\Omega^{(\rho\lambda)}[S]$ and $C_\Omega^{(\rho\lambda)}[A]$ are shown in the left plot in Fig.~\ref{fig:Crl}. From this plot one can conclude that the mixed term $\tilde{\beta}_\Omega^{(\rho\lambda)}$ provides a contribution to the full chromopolarizability at the level of few percent (see the black solid curve) and as such can be neglected. This result should not come as a surprise given that the octets mixing potential (\ref{eq:VAS}) is antisymmetric with respect to the coordinates inversion, ${\bm\rho}\to-{\bm\rho}$ and ${\bm\lambda}\to-{\bm\lambda}$, while we consider the ground-state baryon made of heavy quarks with the wave function symmetric with respect to this coordinates change.

We now turn to the diagonal contributions given in Eqs.~(\ref{br}) and (\ref{bl}) and rewrite them using the same approximation as was used above for the mixed term, \begin{eqnarray}
\tilde{\beta}_\Omega^{(\rho)}\approx\frac{1}{12}\left(
\frac{\braket{\rho^2\sin^2\theta_o}}{\braket{\hat{T}({\bm p}_\rho,{\bm p}_\lambda)}+\braket{V_+({\bm\rho},{\bm\lambda})}-E_\Omega}\right.\nn\\
\left.+\frac{\braket{\rho^2\cos^2\theta_o}}{\braket{\hat{T}({\bm p}_\rho,{\bm p}_\lambda)}+\braket{V_-({\bm\rho},{\bm\lambda})}-E_\Omega}\right),\nn\\
\label{pol6}\\[-2mm]
\tilde{\beta}_\Omega^{(\lambda)}\approx\frac{\zeta^2}{36}\left(
\frac{\braket{\lambda^2\cos^2\theta_o}}{\braket{\hat{T}({\bm p}_\rho,{\bm p}_\lambda)}+\braket{V_+({\bm\rho},{\bm\lambda})}-E_\Omega}\right.\nn\\
\left.+\frac{\braket{\lambda^2\sin^2\theta_o}}{\braket{\hat{T}({\bm p}_\rho,{\bm p}_\lambda)}+\braket{V_-({\bm\rho},{\bm\lambda})}-E_\Omega}\right).\nn
\end{eqnarray}
The result of the direct numerical calculation based on Eq.~(\ref{pol6}) is shown in the right plot of Fig.~\ref{fig:Crl} and compared with both the exact result without octets mixing [for $\theta_o=0$; see Eq.~(\ref{Comegadef})] and a similar approximate result also obtained in neglect of the mixing [see Eq.~(\ref{pol3})]. Thus one can see that the difference between the above three curves is small and can be regarded as a systematic uncertainty. 

Summarizing the results obtained in this appendix one can state that
\be
\beta_\Omega\approx \tilde{\beta}_\Omega^{(\rho)}+\tilde{\beta}_\Omega^{(\lambda)}\approx \beta_\Omega^{(\rho)}+\beta_\Omega^{(\lambda)},
\ee
which justifies working in the zero-mixing approximation in Sec.~\ref{sec:pola}. 

The mixing term $\beta_\Omega^{(\rho\lambda)}$ is then treated as a source of the systematic uncertainty; from the left plot in Fig.~\ref{fig:Crl} one can see that the corresponding contribution $C^{(\rho\lambda)}_\Omega$ does not exceed about 7\% of the total $C_\Omega$.

\end{document}